\documentclass[apj, twocolumn]{emulateapj}

\usepackage{color}
\usepackage[breaklinks]{hyperref}
\usepackage{graphicx}
\usepackage{natbib}
\usepackage{amsmath}
\usepackage{amssymb}
\usepackage{rotating}

\usepackage[math, probs, units, fig, names, comments]{mf}

\begin{document}

\shortauthors{Fouesneau et al.}  
\lefthead{\textsc{Fouesneau et al.}}
\righthead{\textsc{PHAT: Year 1 Clusters Ages and Masses}}

\title{The Panchromatic Hubble Andromeda Treasury V: Ages and Masses of the Year
1 Stellar Clusters
\footnote{Based on observations made with the NASA/ESA Hubble Space Telescope,
obtained from the Data Archive at the Space Telescope Science Institute, which
is operated by the Association of Universities for Research in Astronomy, Inc.,
under NASA contract NAS 5-26555}
}
\author{ Morgan Fouesneau\altaffilmark{1},
	L. Clifton Johnson\altaffilmark{1},
	Daniel R. Weisz\altaffilmark{1,2,13},
	Julianne J. Dalcanton\altaffilmark{1},
	Eric F. Bell\altaffilmark{3},
	Luciana Bianchi\altaffilmark{4}
	Nelson Caldwell\altaffilmark{5},
	Dimitrios A. Gouliermis\altaffilmark{6,7}
	Puragra Guhathakurta\altaffilmark{8},
	Jason Kalirai\altaffilmark{9},
	S{\o}ren S. Larsen\altaffilmark{10},
        Hans-Walter Rix\altaffilmark{7}, 
	Anil C. Seth\altaffilmark{11}, 
	Evan D. Skillman\altaffilmark{12}, and
	Benjamin F. Williams\altaffilmark{1}
       } 
\altaffiltext{1}{Department of Astronomy, University of Washington, Seattle, Washington, USA}
\altaffiltext{2}{Department of Astronomy, University of California at Santa Cruz, 1156 High Street, Santa Cruz, CA, 95064}
\altaffiltext{3}{Department of Astronomy, University of Michigan, 500 Church Street, Ann Arbor, MI 48109, USA}
\altaffiltext{4}{Department of Physics and Astronomy, Johns Hopkins University, Baltimore, MD 21218, USA}
\altaffiltext{5}{Harvard-Smithsonian Center for Astrophysics, 60 Garden Street Cambridge, MA 02138, USA}
\altaffiltext{6}{Institut f\"ur Theoretische Astrophysik, Zentrum f\"ur Astronomie der Universit\"at Heidelberg, Albert-Ueberle-Strasse 2, 69120 Heidelberg, Germany}
\altaffiltext{7}{Max-Planck-Institut f\"ur Astronomie, K\"onigstuhl 17, 69117 Heidelberg, Germany}
\altaffiltext{8}{University of California Observatories/Lick Observatory, University of California, 1156 High Street, Santa Cruz, CA 95064, USA} 
\altaffiltext{9}{Space Telescope Science Institute, 3700 San Martin Drive, Baltimore, MD 21218, USA}
\altaffiltext{10}{Department of Astrophysics, IMAPP, Radboud University Nijmegen, P.O. Box 9010, 6500 GL Nijmegen, The Netherlands}
\altaffiltext{11}{Department of Physics and Astronomy, University of Utah, Salt Lake City, UT 84112, USA}
\altaffiltext{12}{Department of Astronomy, University of Minnesota, 116 Church Street SE, Minneapolis, MN 55455, USA}
\altaffiltext{13}{Hubble Fellow}
\email{mfouesn@astro.washingtonedu}

\begin{abstract} 

We present ages and masses for 601 star clusters in M31 from the analysis of the
six filter integrated light measurements from near ultraviolet to near infrared
wavelengths, made as part of the Panchromatic Hubble Andromeda Treasury (PHAT).
We derive the ages and masses using a probabilistic technique, which accounts
for the effects of stochastic sampling of the stellar initial mass function.
Tests on synthetic data show that this method, in conjunction with the exquisite
sensitivity of the PHAT observations and their broad wavelength baseline,
provides robust age and mass recovery for clusters ranging from $\sim10^2 -
2\times10^6\msun$. 
We find that the cluster age distribution is consistent with being uniform over
the past $100 \Myr$, { which suggests a weak effect of cluster disruption within
M31}.  The age distribution of older ($>100\Myr$) clusters fall towards old ages,
consistent with a power-law decline of index $-1$, { likely from a
combination of fading and disruption of the clusters}.  We find that the mass
distribution of the whole sample can be well-described by a single power-law
with a spectral index of $-1.9 \pm 0.1$ over the range of
$10^3-3\times10^5\msun$.  However, if we subdivide the sample by galactocentric
radius, we find that the age distributions remain unchanged.  However, the mass
spectral index varies significantly, showing best fit values between $-2.2$ and
$-1.8$, with the shallower slope in the highest star formation intensity
regions. We explore the robustness of our study to potential systematics and
conclude that the cluster mass function may vary with respect to environment.

\end{abstract}
\keywords{Galaxies: Individual (M31), Star clusters --- Methods: data analysis,
statistical --- Techniques: photometric}

\maketitle

\section{Introduction} 

It has become clear that a significant fraction of star formation occurs in
stellar clusters. However, deriving galaxy histories from observations of
clusters is complicated by significant uncertainties.  Controversial questions
have been raised regarding cluster properties in various environments as
different analyses could lead to different conclusions.  Claims exists for
interesting variations or trends in cluster colors, their lifetimes as
gravitationally bound objects, and age or mass distributions within or between
galaxies, and the evolution from initial to current cluster mass functions
\citep[\eg,][]{Zepf1993, Kumai1993, Girardi1995, Elmegreen1997, Bastian2011}. 

Such studies rely on our ability to estimate intrinsic properties of stellar
clusters, in particular, their ages and masses. Much observational effort has
therefore been invested in determining the distributions of star cluster ages
and masses \citep{Searle1980, Larsen2000, Billett2002, Hunter2003, Fall2005,
Rafelski2005, Dowell2008, Larsen2009, Chandar2010, Bastian2012}, and using the
resulting data to determine the dominant mechanisms of cluster formation and
disruption \citep{Kroupa2002,Boutloukos2003, Lamers2005, Whitmore2007,
Parmentier2008, Fall2009, Elmegreen2010, Converse2011}.  However, most of the
existing work deals with observations of relatively massive clusters (a few
$10^4-10^5\msun$), which are the least affected by disruption processes and thus
are the most stable in various environments.  As a result, divergences
between disruption models \citep[\eg][among many others]{Baumgardt2003,
Fall2009} are still subject to debate in the literature
\citep[\eg][most recently in M83]{Whitmore2011, SilvaVilla2014arXiv}. Only a
few studies have been able to probe the smaller clusters that are the most
sensitive to environmental effects (mainly in the Galaxy, \eg,
\citealt{Borissova2011}, or the Magellanic Clouds, \eg, \citealt{Popescu2012}). 

The Panchromatic Hubble Andromeda Treasury (PHAT; \citealt{Dalcanton2012a}) is
an ongoing multi-cycle Hubble Space Telescope (HST) program that is ideal for
studying stellar clusters in M31.  The survey imaged one-third of the M31 disk
at high spatial resolution with wavelength coverage from the ultraviolet through
the near-infrared.  The sensitivity of the latest HST instruments allow us to
detect clusters in M31 down to a regime in which cluster luminosities overlap
those of individual bright stars, and hence to very low masses
\citep{Johnson2012a}.  This survey spans a wide range of environments in both
star formation intensities and gas densities. This diversity is an advantage for
addressing how environment affects cluster formation.

The PHAT survey has already significantly increased the number of clusters known
in M31. \citet{Johnson2012a} identified $601$ stellar clusters using the first
quarter of the total PHAT coverage. This new cluster catalog contains more than
a factor of four increase in the number of known clusters within the survey
area.  Moreover, the uniform photometric coverage from the UV to near-IR allows
accurate age-dating of the clusters.  Even this preliminary sample breaks new
ground for studying clusters outside the Milky Way and the Magellanic Clouds,
probing  about two orders of magnitudes fainter in the luminosity function
\citep[their figure 11]{Johnson2012a}.

This paper is part of a series utilizing the PHAT dataset for studies of stellar
clusters.  \citet{Johnson2012a} presented the first installment of a HST-based
cluster catalog, which serves as the basis for an extensive study of Andromeda's
cluster population.  In this paper, we focus on the determination of ages and
masses of the first year sample, looking forward to the final product of this
four year Treasury program. Our estimates of the properties of the clusters are
derived from integrated photometry in six broad bands and we especially focus
our attention on the characterization of the lowest-mass clusters.  Additional
studies, including analysis of structural parameters, resolved star content, and
integrated spectroscopy of the cluster sample will follow in subsequent work.

\medskip

This paper is organized as follows.  \S \ref{sec:data} presents the cluster
sample and the key elements of their photometry. \S \ref{sec:Analysis} describes
the analysis and the cluster models used to derive the properties of the
clusters, and briefly highlights the possible artifacts of the method using
synthetic data.  \S \ref{sec:yr1properties} describes our results for the entire
sample and for individual regions across M31.  Finally, we discuss those results
in \S \ref{sec:discussion} before drawing our conclusions.

\section{Observations \& Cluster catalog} 
\label{sec:data}

For this paper, we use the list of $601$ high-probability cluster candidates
from the \citet{Johnson2012a} Year 1 catalog, which contains integrated
photometry through six broad-band filters from the UV to the near infrared:
F275W ($UV$), F336W ($U$), F475W ($g$), F814W ($I$), F110W ($J$), F160W ($H$).
Clusters were detected by-eye, primarily based on the F475W images, and visually
classified based on their sizes, shapes, and concentrations as explained in
\citet{Johnson2012a}.

\begin{figure*}
  \centering
   \includegraphics[width=0.9\textwidth]{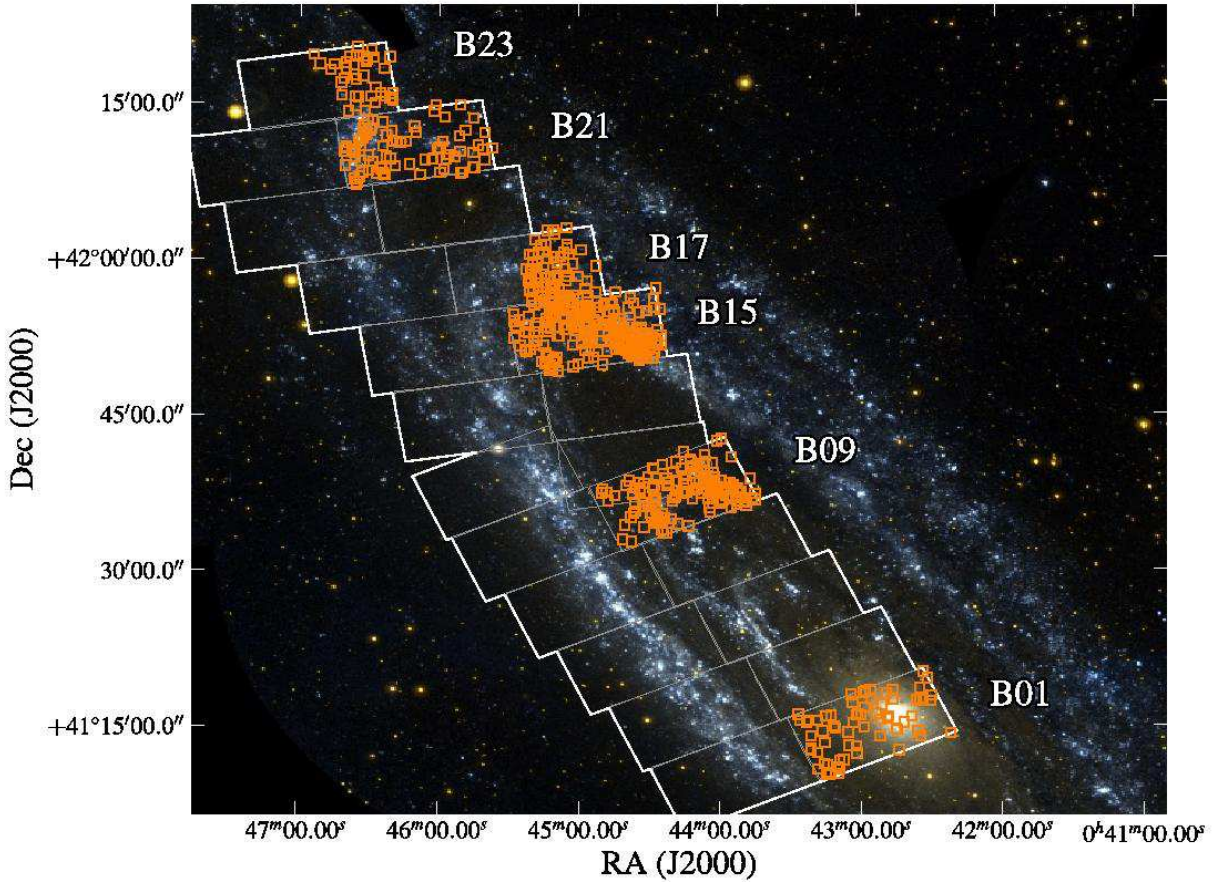}
   \caption{ GALEX FUV/NUV composite image showing the positions of the clusters
   (orange squares) in the PHAT Year 1 Catalog. The white outer contour
   illustrates the footprint of the final survey, subdivided into rectangular
   ``bricks'' ($1.5\times2.7\kpc$), shown in grey. Labels correspond to the names
   of the regions mentioned in \S\,\ref{sec:data}. }
  \label{fig:sources_pos}
\end{figure*}

The Year 1 catalog is sub-divided into regions called ``bricks'', following the
survey observation strategy described in \citet{Dalcanton2012a}. The cluster
catalog includes four full bricks (designated B01, B09, B15, and B21; numbers
increase with increasing galactocentric radius) as well as the western halves of
two additional bricks (B17W and B23W).  For simplicity,  we group B17W with B15
and B23W with B21, respectively, for the remainder of this work, since both
pairs form contiguous regions. These data sample locations along the major axis
of M31, covering the bulge (B01) and regions at at $\sim6$, $10$, and $15\kpc$
from the center of the galaxy (B09, B15, and B21, respectively). With the
exception of the bulge-dominated brick, B01, the remaining ones target regions
of high star formation, located on the star forming ring (B15) or on spiral arms
(B09, B21). Of the three disk fields, B15 samples the highest star formation
intensity, found in the ``$10\kpc$-ring''.  The color image in
Fig.~\ref{fig:sources_pos} illustrates the positions of the clusters in the
catalog in the different observed regions, relative to the expected final
coverage of the survey.

Measurement details are described in \S\,4 of \citet{Johnson2012a}, we briefly
summarize the different steps in the following.  We measured instrumental
magnitudes for the clusters using aperture photometry.  Measurements were
converted into the VEGA magnitude system \citep[see][\S 4.1.1 for
details]{Johnson2012a}. We do not perform passband conversions and instead work
with the native HST filters.  For each object, circular aperture photometry
produced integrated flux values, and aperture radii vary from cluster to cluster
between $0.5\arcsec$ ($1.9\pc$) and $6\arcsec$ ($22.7\pc$).  Measurement
uncertainties are typically lower than $0.2$\,mag, for which we include
background measurement uncertainties. Aperture corrections were based on their
half-light radius in the F475W images, assuming flat radial color profiles in
the outer parts of the clusters.  Typical correction values are $\sim$
$0.1$\,mag. 
We assume a distance to M31 of $785 \kpc$ \citep{McConnachie2005}, which
corresponds to a distance modulus of $m-M = 24.47$ mag.

Assessing the completeness of cluster samples is a challenge on its own.  The
true completeness of the sample as a whole is a complicated function of cluster
luminosity, size, and location within M31.  The completeness of the sample was
assessed by conducting artificial cluster tests. Briefly, we used synthetic
clusters that span the range of properties expected for the cluster sample.
Specifically, we sampled ages from 4 million to 10 billion years and masses from
$10^2$ to $10^5\msun$, and cluster half-light radii (assuming \citealt{King1962}
profiles) of $1 - 7\pc$  \citep[\S 3.1 of][especially their
Fig.\,6]{Johnson2012a}.  On average, the $50$\% completeness limits in the F475W
filter are M$_{F475W} = -3.8$, $-3.5$, $-2.8$, $-2.2$ for B01, B09, B15, and
B21, respectively.

\section{Deriving Cluster properties}
\label{sec:Analysis}

In this section, we derive cluster ages and masses using the probabilistic
method developed in \citet{Fouesneau2010}.  The method is based on a large
collection of Monte-Carlo simulations of individual clusters. These simulations
take full account of stochastic sampling of the stellar mass distribution,
allowing robust Bayesian fitting to the observed colors of the observed
clusters, even for the low mass regime.

\subsection{Population models}
\label{sec:models}
Synthetic spectral energy distributions (SEDs) of clusters are constructed with
the population synthesis code {\tt P\'egase.2n} (Fouesneau et al. in prep.),
which is derived from {\tt P\'egase} \citep{Fioc1997}. As in the original
population synthesis code, the underlying stellar evolution tracks are those of
the Padova group \citep{Bressan1993, Bertelli1994}, with a simple extension
through the thermally pulsating AGB based on the prescriptions of
\citet{Groenewegen1993}. The input stellar spectra are taken from the library of
\citet{Lejeune1997}. The stellar initial mass function (IMF) is taken from
\citet{Kroupa1993}, and extends from $0.1$ to $120\msun$.  Nebular emission
(lines and continuum) are computed as in \citet{Fioc1997} and are included in
the calculated spectra and broad band fluxes under the assumption that no
ionizing photons escape. The photometry for the synthetic clusters is computed
using the response curves of the PHAT HST/ACS and HST/WFC3 filters. A reference
spectrum of Vega provides zero magnitude fluxes \citep{Bohlin2007}. 

The code uses Monte-Carlo (MC) methods to populate the stellar mass function
(SMF) with a finite number of stars. The simulations draw an explicit
number of stars instead of a target mass which avoids the possible biases linked
to the latter case \citep{Kroupa2013}. As a result, models explicitly account
for stochastic variations in the stellar content of clusters.  We have extended
the simulated cluster set to lower masses than available in
\citet{Fouesneau2012}, and the coverage now ranges from $\sim50$ to
$5\times10^5\msun$, and ages from $1\Myr$ to $20\Gyr$.  With a few $ \times
10^8$ individual models, the collection of synthetic clusters is large enough to
include all reasonably likely cluster properties.

We also include a transition to continuous models above $2\times10^5\msun$ (\ie,
models assuming a continuously populated stellar mass function), motivated by
the presence of about $\sim 30$ massive globular clusters in the PHAT catalog
\citep[$>10^5\msun$,][]{Caldwell2011}.  For computational reasons discussed in
\citet{Fouesneau2012}, the mass distribution in the collection of models follows
a power law of index $-1$.  The ages of the synthetic clusters are drawn from a
power law distribution with index $-1$ (equal numbers per logarithmic bin),
rounded to integer multiples of $10^6 \yr$.   The extinction, $A_V$, is allowed
to vary uniformly from $0$ to $3$ magnitudes with a fixed $R_V$ of $3.1$
assuming the standard extinction law of \citet{Cardelli1989}. 

Based on HII region abundances \citep[\eg,][]{Zurita2012}, we expect young
clusters in M31 to have approximatively solar metallicity. Thus, we fix the
metallicity, $Z$, to $0.02$ (solar) for the discrete part of the collection.  In
contrast, globular clusters are known to have lower metallicities
\citep[\eg,][]{Caldwell2011,Cezario2012}, and thus the metallicity of the
continuous models for massive clusters is allowed to vary (equiprobably) between
the fixed values of $[0.004, 0.008, 0.02, 0.05]$, corresponding approximately to
Small and Large Magellanic Clouds, Solar, and super-Solar metallicities,
respectively. 

Individual cluster estimates are collected into Table \ref{tab:catalog} in
Appendix\,\ref{sec:appendix_catalog}.

\subsection{Analysis Method}
\label{sec:method}
The probabilistic method developed in \citet{Fouesneau2010} calculates posterior
probability distributions in the age-mass-extinction space (marginalized over
the metallicity dimension), using multi-wavelength photometric observations and
the large collection of Monte-Carlo simulations of clusters of finite stellar
masses, described above.  As in all Bayesian inference approaches, the results
are stated in probabilistic terms, and they depend on \apriori~probability
distributions of some model parameters (priors). 

In our context, the probability for one cluster to have a specific age, mass,
extinction, and metallicity, given a set of photometric observations is:
\begin{equation}
  \proba{\T \given \data, \err} \propto \proba{ \data \given \T, \err} \times \proba{\T},
  \label{eq:bayesian}
\end{equation}
where the $\data$ and $\err$ are the $6$ band photometric measurements and
associated uncertainties, $\T$ is the ensemble of parameter values (\ie,
age-mass-extinction-metallicity), $\proba{\T}$ are the priors on the values of
the parameters $\T$, and $\proba{\data \given \T, \err}$ is the likelihood of
measuring an energy distribution for a given set of $\T$. 

\paragraph{Likelihood}
We adopt a normal based-likelihood function, assuming the photometric errors are
Gaussian and independent flux measurements:
\begin{eqnarray}
  \proba{\data \given \T} =  
  \prod_{k=1}^6 \frac{1}{\sqrt{2\pi}\sigma_k} \exp\bra{-\frac{\pa{d_k - \hat{d_k}(\T)}^2}{2\,\sigma_k^2}}
  \label{eq:likelihood}
\end{eqnarray}
in which $k$ indicates properties for the $k$-th filter, and $\hat{d_k}(\T)$ the
predicted flux in the filter $k$ for the given set of $\T$.

\paragraph{Priors}
The posterior probability distribution, $\proba{\T \given \data}$, depends on
the prior constraints on the values of the parameters $\T$. This prior
information translates the age and mass distributions of the synthetic clusters,
together with the values allowed for extinction. As described in
\S\,\ref{sec:models}, our priors can be explicitly expressed as an independent
combination of age, mass, and extinction functions:
\begin{eqnarray}
  \proba{\T} \propto (M/\msun) ^ {-1} \times (A/\yr) ^ {-1},
  \label{eq:priors}
\end{eqnarray}
corresponding to independent power-laws in age and mass, and implicitly uniform
in $A_V$, respectively.  The completeness limits of the observations are not
used in any manner during the determination of individual ages and masses.  The
power-law distributions we have adopted mimic two major qualitative trends seen
in star-forming galaxies: low mass clusters are more numerous than high mass
clusters, and, because of a variety of efficient disruption mechanisms, young
clusters are more numerous than old ones. As demonstrated in
\citet{Fouesneau2012}, the priors do not globally dominate the resulting
behavior of $\proba{\T \given \data, \err}$ (Eqn.\,\ref{eq:bayesian}). A study
based on the full survey will further optimize the prior distributions, in
particular age-mass, to derive cluster disruption efficiencies across the
covered area of M31.

\begin{figure*}
  \centering
  \includegraphics[width=0.8\textwidth]{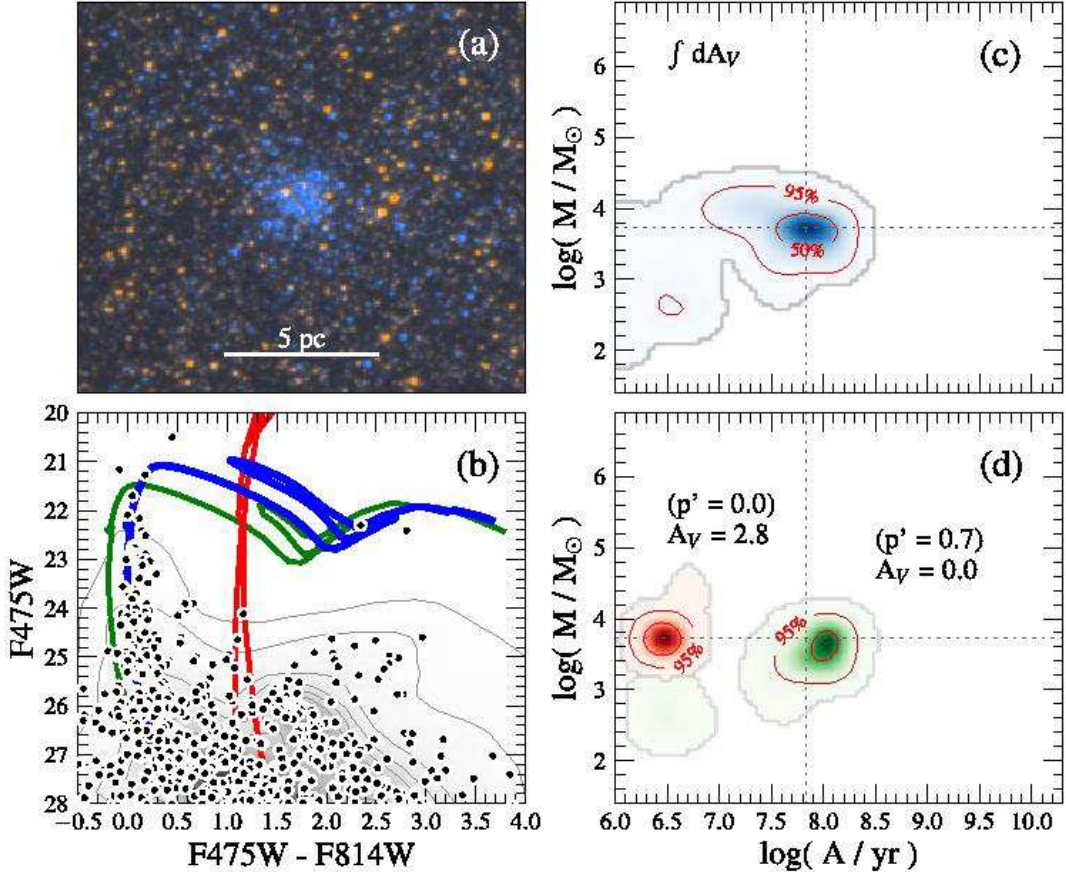}
  \caption{ Illustration of the estimated age-mass posterior distribution for a
  given cluster in the PHAT sample and comparison with its color-magnitude
  diagram (\latin{c.f.}, \S\,\ref{sec:method}). From top to bottom, and left to
  right:
  {\bf (a)} Composite image of optical HST observations of a typical cluster in
  the sample;
  {\bf (b)}  Color Magnitude diagram of the cluster using PHAT photometry and
  isochrones based on the color-coded best fits shown in panels (c) and (d).
  Black dots represent the colors and magnitudes of individual stars in the
  photometric aperture centered on the cluster. The grey density and contours
  show the background population distribution in the color-magnitude space,
  allowing a quantitative estimate of the stellar membership; 
  {\bf (c):} This panel represents the
  age$-$mass posterior distribution of this particular cluster while accounting for the
  full range of allowed extinction values (marginalized over \Av).  Contours
  follow the same definition as in the lower right panel.
  We obtain a complex distribution in which the most probable value is the
  triplet: $\set{\Av: 0.3 \,;\, \log(\mathrm{A}):7.8\,;\,\log(\mathrm{M}):3.7}$,
  of which a projection is indicated by the dotted lines on the right hand side
  panels;
  {\bf (d)} illustrative representation of posterior distributions for two
  slices in the extinction dimension,  $\Av=0$ in green and $\Av=2.8$ in red (distributions are thus
  independently normalized). The two most likely age-mass-av triplets lead to the
  green and red isochrones on the bottom-left panel. For each posterior, contours
  indicate limits at which a given fraction of the distribution is enclosed: red
  contours indicate $50$\% and $95$\%, and a grey contour $99.9$\% and $p\prime$
  the respective probability ratio to the best fit.
  }
  \label{fig:pdf_example}
\end{figure*}

An illustrative example of the method for a typical cluster is given
Fig.\,\ref{fig:pdf_example}. A composite image of the cluster is shown in panel
(a). We show the cluster color magnitude diagram (CMD) in panel (b), on which
the gray density map in the background represent the possible contamination by
the field stars. On this CMD, we overlaid a blue isochrone corresponding to the
best fit value from the integrated photometry  analysis accounting for
age-mass-extinction correlations.  We represent the correlation between age and
mass by the joined age-mass probability distribution function (PDF) in panel (c)
of Fig.\,\ref{fig:pdf_example}, marginalized over the extinction parameter,
$A_V$.

In panel (d) of Figure\,\ref{fig:pdf_example}, we demonstrate the possible
effect of $A_V$ by considering the joined age-mass PDF at $2$ distinct slices of
extinction, before integrating over all $A_V$ as in panel (c). We first consider
a slice a zero extinction, which is close to the best fit extinction of $0.3$.
We represent this PDF in green on panel (d) and the corresponding green
isochrone on panel (b). We also show a slice at high extinction ($A_V=2.0$)
represented in red with also a corresponding red isochrone on panel (b).  This
latter slice contains a very small fraction of the complete PDF, but has been
renormalized to allow it to be visible in panel (d).

Figure\,\ref{fig:pdf_example} also demonstrates the subtleties in interpreting
PDFs. First, the best value does not always correspond to the peak in a
marginalized PDF. The dotted lines indicate the position of the best fit in the
age-mass-$\Av$ space, \ie, of the triplet that maximizes the posterior
distribution in the full parameter space.  Although very close to the peak of
the age-mass distribution in panel (c), the best fit suggests a slightly younger
and more massive cluster.  Second, distributions are complex. The strong
correlation between all the different parameters leads to distributions far from
Gaussian and sometimes multi-modal. 

From this analysis of the integrated light of this particular cluster in the
PHAT sample, we obtain a qualitatively good fit, according to the CMD locus of
the blue isochrone on panel (b). This panel suggests that the fit indeed
accounts for the presence of 2 bright red stars, a significant number of bright
main sequence stars, and a reddened main sequence. We also expect that the
extended distribution towards younger ages (see Fig\,\ref{fig:pdf_example}.d)
captures the presence of the two brightest stars on the CMD. 

\subsection{Catalog}

Appendix \ref{sec:appendix_catalog} gives a table of individual estimates
for each cluster of the sample, for the discrete models (including continuous
extension at the higher mass end).  The quoted values in this table are the
age-mass-extinction triplets that maximize the posterior probability over the
full parameter space. Quoted uncertainties are based on the percentiles of the
posterior distribution. As we mentioned above, only the full posterior
distribution keeps the complexity of the information. While PDFs from this
preliminary study are available upon request, full PDFs will be made available
with the cluster catalog when the survey will be completed.

We characterized the potential artifacts and biases in the determination of
cluster ages, masses, and extinctions using samples from the synthetic cluster
collection.  Briefly, for a sample of synthetic clusters spanning the full range
of cluster ages, masses and extinctions, we generated $6$-filter photometry for
one sample of synthetic clusters, and perturbed the ``measured magnitudes''
according to uncertainties distributed as in the actual PHAT cluster data.  We
then re-derived the properties for this synthetic set of clusters, and compare
the recovered values to the input values.

Overall, we do not find significant age or mass biases from the analysis of
synthetic data.  The dispersions we obtain are $\sim$0.14 dex in mass and
$\sim$0.18 dex in age.  Those dispersions are consistent with the scatter
expected on the basis of the derived PDFs. Based on these tests, we will not
venture to interpret features smaller than 0.2 dex in either age or mass
(conservatively, since the test dataset corresponds to perfectly modeled data).
To reflect these limitations, we bin the posterior probability distributions to
0.2 dex in all the subsequent figures.  

We find that large errors in age occur for a few percent of the different
realizations of the synthetic clusters, mostly at ages of one or a few Gyr.
These errors are due to the age--extinction degeneracy, coupled with the
addition of the metallicity as a new parameter for the high mass regime. However
these failures can easily be detected through a visual inspection of the
color-magnitude diagrams (CMDs). Therefore, we used panel (b) in
Fig.\,\ref{fig:pdf_example} as our baseline to visually inspected the CMDs of
each observed cluster and added a caution flag in Table \ref{tab:catalog} when
we estimate a potential failure of the fit. These cases include $\sim60$
clusters, mainly in the bulge.  In further studies, we will include independent
determinations of the extinction and/or metallicity when available, providing a
better set of initial priors.

\subsection{Comparison with color-magnitude diagram analysis}
\label{sec:cmd}
\begin{figure}
  \centering
    \includegraphics[width=\columnwidth]{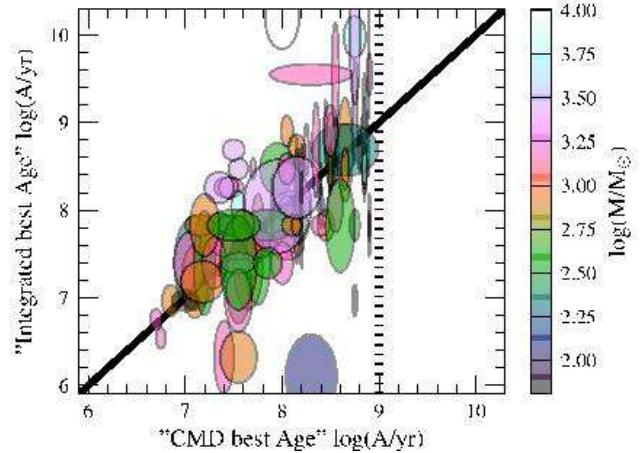}
  \caption{ 
  Comparison of age estimates from this present work with color-magnitude
  based estimates for a 100 clusters with visible main sequence. The horizontal
  axis shows the best age estimates from CMD analysis using MATCH
  \citep{Dolphin2002b}, and the vertical axis represents the best fit values
  from this present study. Each ellipse represents the uncertainty ellipse of
  the cluster point, and the colors encode the mass of the cluster.  The dotted
  line is the limit in age imposed during the fitting procedure of the CMDs and
  the solid line indicates the identity function. This figure shows a broad
  agreement of both methods within their respective uncertainties.
    }
  \label{fig:age_comp}
\end{figure}

{
In this present study, we estimate ages and masses by analyzing the integrated
photometry measurements of the clusters. We could argue that such method is
uncertain and may not give accurate estimates, especially when the energy
distributions of the clusters are deeply affected by stochasticity and stars in
rare evolutionary phases with unusual colors.  Although limited to nearby
galaxies, where clusters can be resolved into stars, the analysis of CMDs offers
more robust estimates.

Taking advantage of the high-resolution HST imaging, PHAT provides individual
star photometry for a significant number of clusters.  To verify the ages and
masses obtained from integrated light, we conducted a CMD analysis of $100$
clusters. This CMD analysis will be presented in detail in Beerman et al. {\it
in prep}; we briefly summarize the method here.  The analysis method is based on
\citet{Dolphin2002b}, which optimizes the likelihood of the observed CMDs with
theoretical ones created from stellar evolution tracks, or isochrones, for a
variety of ages, mass functions, binary fractions, metallicity, etc., including
all observed errors and contamination from a background field population.  We
selected the sample of $100$ clusters for the CMD analysis from various regions
of the survey, selecting only clusters with a visible main sequence to ensure
the accuracy of the recovered ages.  we include clusters from Because of these
choices, we restricted the CMD models to have ages less than $10^9\yr$.

Figure\,\ref{fig:age_comp} presents a one-to-one comparison of age estimates
derived from integrated photometry (\ie, this present study) with estimates
derived from the CMD analysis. Ellipses are located centered on the best fit
values and their sizes encode the uncertainties from the CMD and integrated
light analyses, respectively.  While both methods have their own caveats and
failure regimes, in general, the  CMD fits have smaller uncertainties.  Our
choice to limit fits to ages younger than $10^9\yr$, is likely to be responsible
for the few massive and old outliers.  Estimated ages are in agreement within
the uncertainties. The dispersion around the 1:1 line is symmetric, suggesting
that integrated light estimates do not present any significant bias toward
younger or older estimates relative to the CMD derived estimates.}

\section{Year\,1 PHAT Clusters properties}
\label{sec:yr1properties}

\begin{figure*}
  \centering
    \includegraphics[width=1.9\columnwidth]{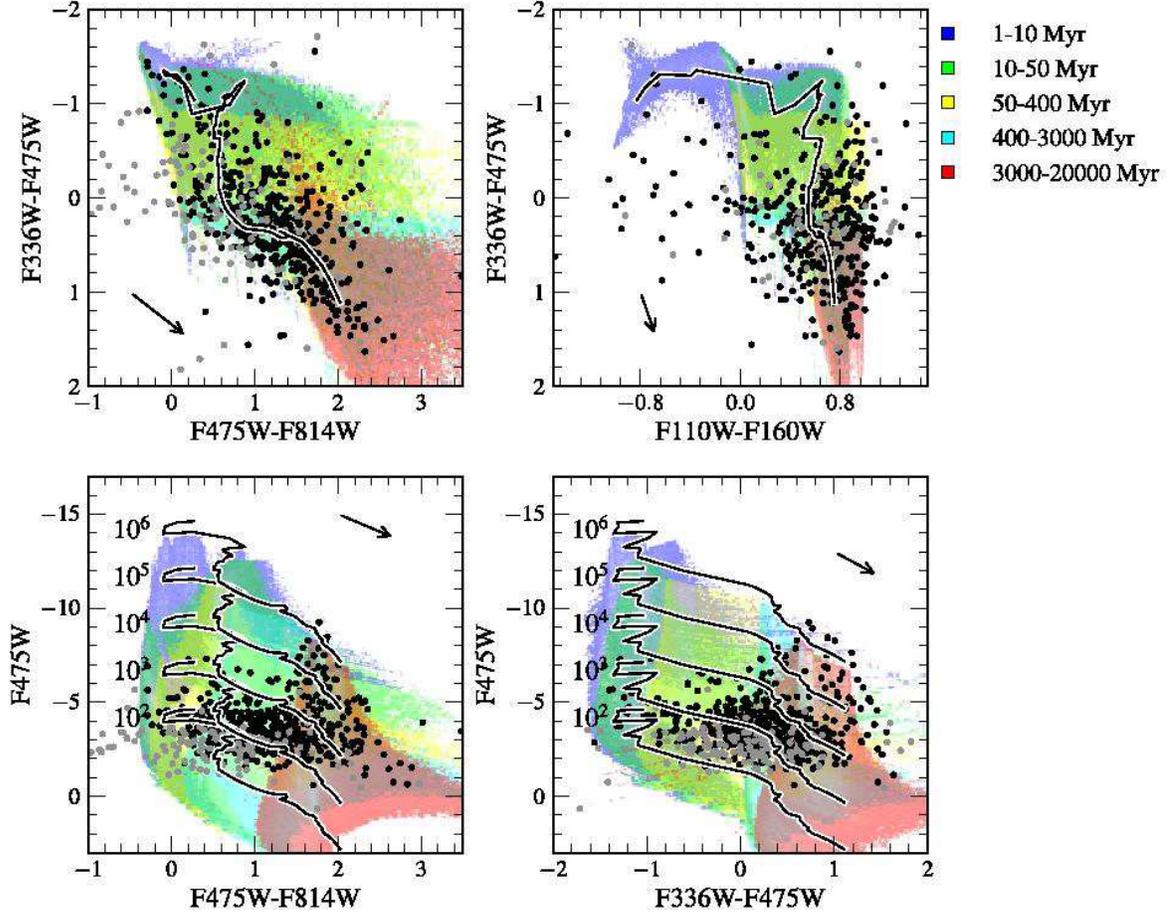}
  \caption{Color-color (top row) and color-magnitude (bottom row) location of
  star clusters with different ages and masses.   The black and gray points
  shows the measured colors and absolute magnitudes of the observed cluster
  sample.  The black points represent optical color uncertainties $\le$ $0.4$
  magnitudes and the gray points uncertainties $>$ $0.4$ magnitudes.
  Unreddened model cluster locations are shown as points whose color encodes age
  (see right hand side), drawn from the model population described in
  \S\,\ref{sec:models}.  The solid lines show the predictions from
  ``continuous'' population synthesis models for clusters masses of $10^2$,
  $10^3$, $10^4$, $10^5$, and $10^6\msun$, respectively. Colored dots represent
  50\% of the discrete clusters available in the Monte-Carlo collection. As the
  models are not reddened in any panel, but the data presumably are, the
  extinction vector of $A_V=1$ is shown for reference. Magnitudes are in the
  Vega system, in the PHAT ACS and WFC3 filters.  The bottom left panel suggests
  a cluster completeness limit of F475W $> -3$. }
  \label{fig:colormags}
\end{figure*}

The current PHAT cluster catalog focuses on the bulge and three major star
forming regions. In this section, we thus look at the first glimpse of what can
be expected from the full balance of the PHAT stellar cluster survey, which will
include many more clusters and will sample a wider range of environment.

\subsection{Global picture}
\label{sec:yr1properties_globalpic}

In Figure~\ref{fig:colormags}, we compare the {loci} of the observations with a
set of unreddened discrete models, in four projections of color-magnitude space.
The data are shown together with half of the unreddened models described in
\S\,\ref{sec:models}, which will be used to assign age, mass, and extinction
estimates to each individual cluster.  These panels illustrate the
dispersions in color and flux that results from the stochasticity inherent to
the discrete nature of the IMF. Models cover broad regions of the diagrams and
complex overlap between ages exist and may not be easily visible at first sight.
This complexity will only increases with the inclusion of reddened models.
Moreover, even though the models plotted in Figure~\ref{fig:colormags} do not
include reddening, the majority of the observed clusters lie well within the
regions covered by the synthetic clusters.  In contrast, {continuous} population
synthesis models (shown by solid lines) are unable to reproduce some of the
observations, even when possible reddening is considered.  For example, the top
left panel of Fig.~\ref{fig:colormags} shows a significant fraction of robust
measurements (black dots) lying on the left side of the { continuous} age
sequence, \ie, the solid line.  Allowing for extinction will not help the
{continuous} models to predict such colors. These colors correspond to those
expected for relatively low-mass clusters ($<$ a few $10^4\msun$) that lack any
post-main sequence stars.   

However, there are also observations even bluer than any models in the
F475W$-$F814W color, lying outside both models range of predicted colors and
fluxes. We have colored all points with large photometric uncertainties
($>0.4$\,mag) with grey, which shows that these outliers are likely to be due to
photometric errors. The UV in particular, F275W and F336W, may still be
adversely affected by cosmic ray artifacts as cautioned in
\citet{Johnson2012a}. 

The locations of the clusters in color and magnitude space draw the unsurprising
picture of a cluster population spanning a wide range of age and mass. The
entire length of the age sequence seems to be populated and the mass range
suggests the presence of a significant number of clusters with masses well below
$10^3\msun$.  The two bottom panels reveal a general trend that the most
luminous clusters in this sample have old ages; these very old clusters fall
above the $10^5\msun$ curve, and are very likely to be old globular clusters.
The typical completeness limit of $-3$ in the F475W band is manifested as the
lower limit to the data in the bottom left panel.

\begin{figure*}
  \centering
    \includegraphics[width=0.9\columnwidth]{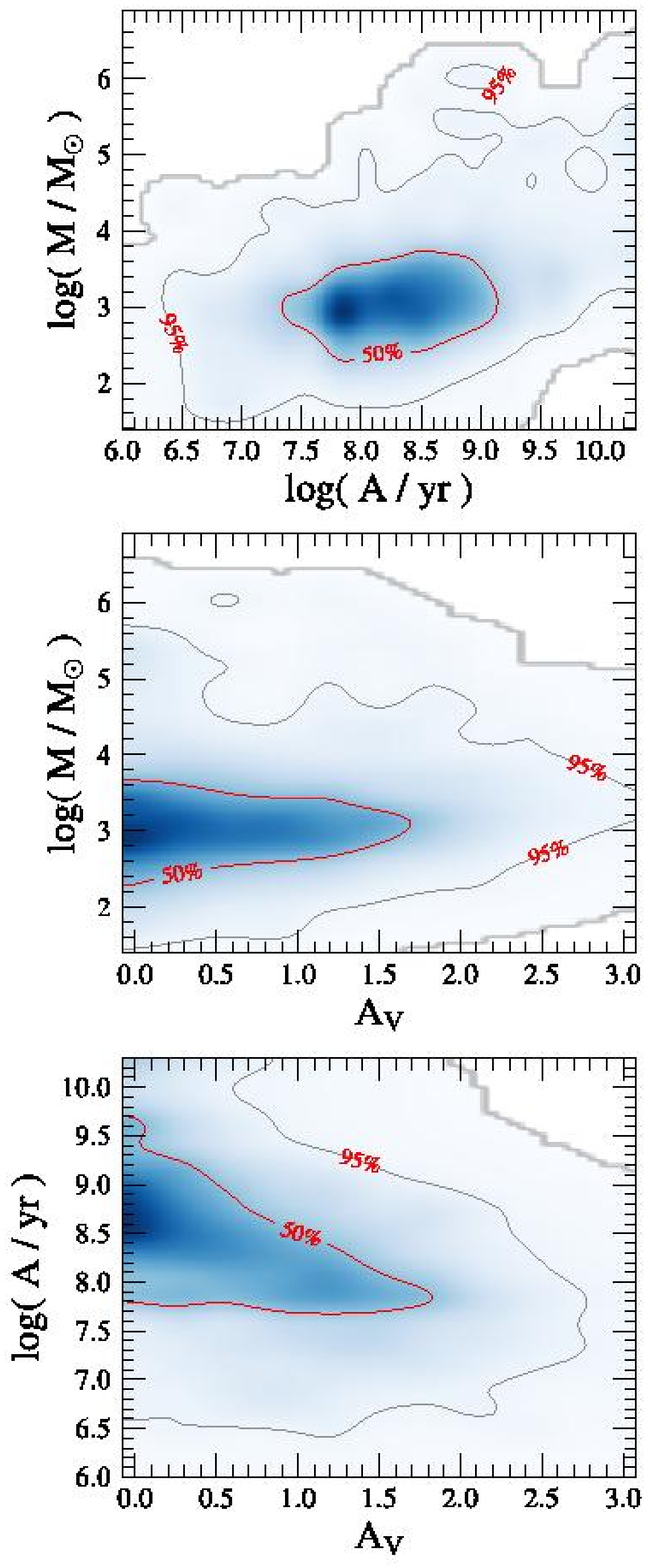}
    \includegraphics[width=0.9\columnwidth]{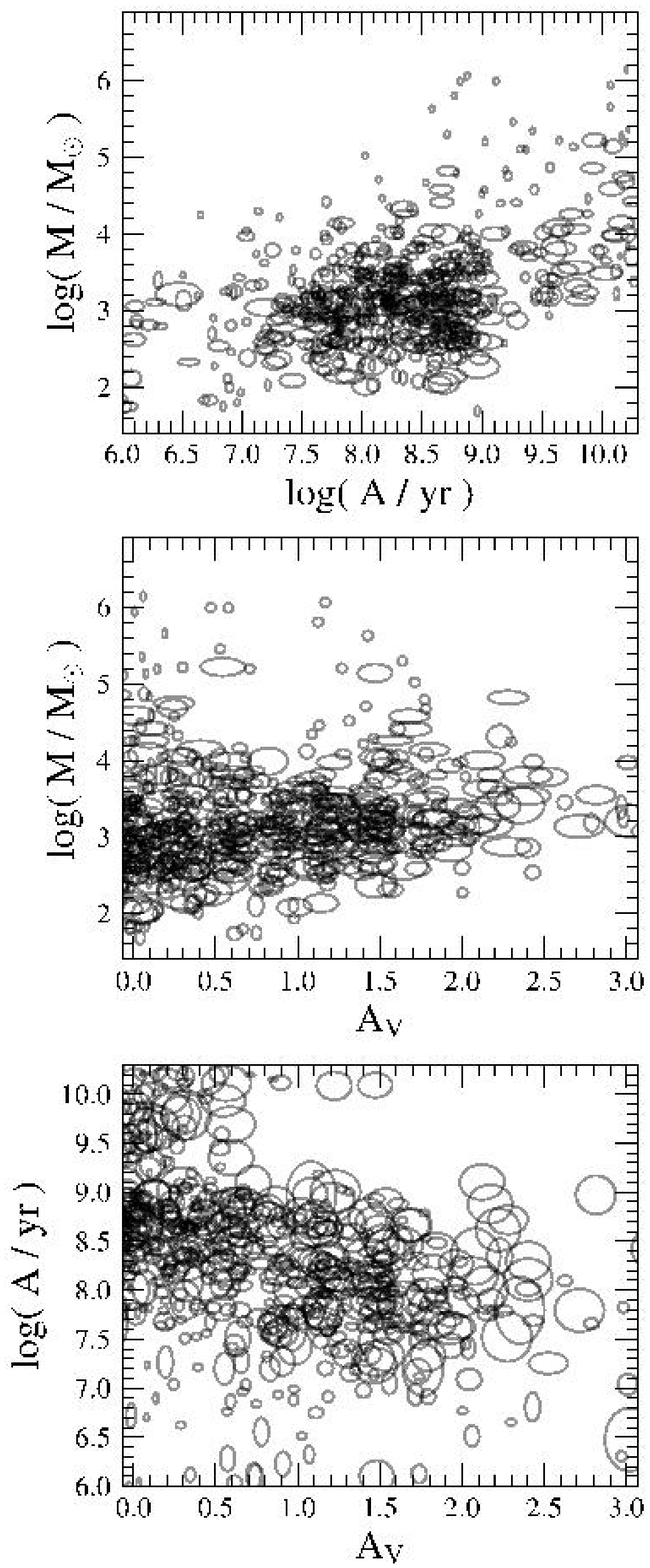}
  \caption{Ensemble probability distributions for age (``A''), mass (``M'') and
  extinction (``\Av'') of all the clusters.
  Each panel on the left represents a distribution marginalized over the third parameter and
  color coded according to the scale on the right side of each panel.  Red
  contours represent the 5\%, 50\%, and 95\% quantiles of the distributions in their
  respective 2D-space. The gray contour represents 99.9\% of the joint distribution.
  Panels on the right represent the locations of the best estimates, where each
  ellipse encodes the relative uncertainties of the individual clusters (an
  arbitrary normalization factor was applied for clarity).
  }
  \label{fig:dist}
\end{figure*}

Figure\,\ref{fig:dist} shows the age-mass-extinction distributions
resulting from the analysis described in \S \ref{sec:Analysis}. Each panel on
the left hand side represents the joint probability distribution of two of the
variables, after marginalizing over the third.  The right panels indicate the
best fit values and their {\it relative} accuracy; the ellipses are arbitrarily
normalized to reduce the clutter in the figures.  The derived ages are
distributed between a few Myr and about $10 \Gyr$, with the few massive
candidates appearing only for ages older than $5 \Gyr$, as expected for the
globular cluster population.  The derived masses range from a little above the
lower limit of our model catalog ($50\msun$) to a bit more than $10^6\msun$.

The significant number of clusters found to have masses below $10^4\msun$ is a
strong argument against using traditional continuous models that ignore the
discrete nature of the IMF. Multiple studies have shown the impact of
stochasticity on the determination of ages and masses of clusters
\citep[\eg][]{Barbaro1977, Bruzual2003, Cervino2004, MaizAppelaniz2009,
Piskunov2009, Fouesneau2010, Popescu2010a}. In particular they identify
artifacts that translate into unphysical overdensities at particular ages for
young and old clusters, while underestimating the number of moderate age
clusters. These artifacts of continuous models are not present in our analysis
(right panels of Fig.\,\ref{fig:dist}), for instance we do not predict an
accumulation of clusters at young ages and instead a relatively smooth age
distribution.

Figure\,\ref{fig:dist} shows almost no old and low mass clusters, nor
highly reddened massive old clusters. At old ages, we do not expect many highly
reddened clusters to be present, since they should have drifted far from their
birth sites.  However, even unreddened old clusters are unlikely to be detected
unless they were very massive, given that less massive clusters would have faded
below the detection limit of the sample.  Cluster disruption processes further
lower the number of old clusters.  Both arguments explain the lack of old,
low-mass clusters in the top panel of Fig.\,\ref{fig:dist}.

There are poor constraints on the number of clusters at the high-mass end,
mainly due to their low birthrates. If we suppose the canonical power-law with
a -2 index, a cluster of $10^5\msun$ is $100$ times less likely to form than a
cluster of $10^4\msun$.  Although a large galaxy like M31 may have formed a few
dozen clusters above $10^5\msun$, they are intrinsically rare. Given that
we expect to observe all the young massive clusters in the galaxy, the lack of
clusters younger than $50\Myr$ compared to older ages ($>1\Gyr$) in the age-mass
distribution suggests that M31 may have been more efficient at producing massive
clusters in the past.  We discuss the decrease in the rate of massive cluster
production in more detail in \S \ref{sec:yr1properties_ages}.

The bottom panel of Fig.\,\ref{fig:dist} shows that very young clusters come
with a large range of extinction values, as has been seen in many star forming
galaxies \citep[\eg,][]{Whitmore2002, Kim2012}.  However, at ages older than
$10^8 \yr$, clusters with more than one magnitude of extinction become rare.
Between $10^7$ and $10^8\yr$, the absence of reddened clusters most likely
reflects a real lack of highly reddened objects, rather than selection effects,
given that the detection limits of the data would allow us to detect $5 \times
10^7 \yr$ old clusters with up to more than $2$ magnitudes of extinction, for
masses above $\log(M/M_\odot) \approx 4$.

\begin{figure}
  \centering
    \includegraphics[width=0.9\columnwidth]{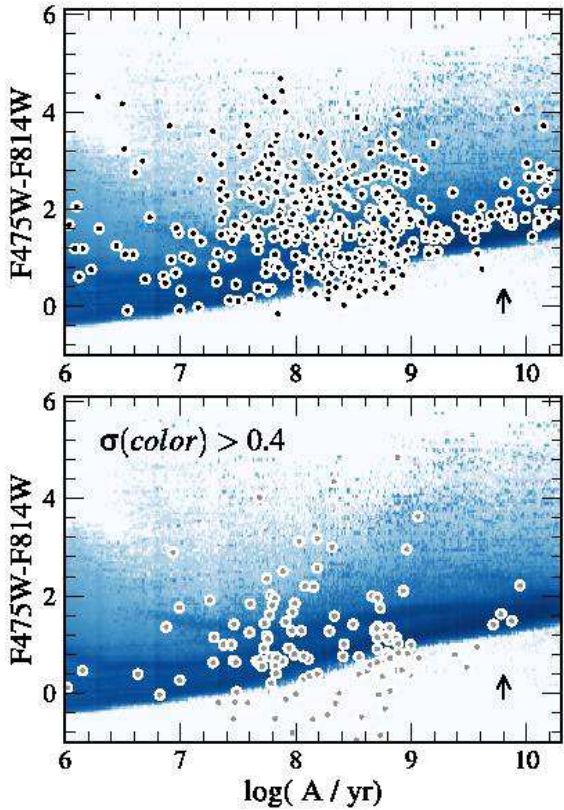}
  \caption{ Comparison of the cluster optical colors with the estimated ages
  from our study.   The density map in the background represents the sample of
  (unreddened) models as shown in Fig.\,\ref{fig:colormags}. One magnitude of
  extinction $A_V$ translates into a variation of 0.69 magnitude in the optical
  color indicated by the arrow in the bottom right corner.  Each point on this
  figure represents a cluster from our catalog that we corrected for reddening
  by applying their respective best fit $A_V$ values.  The top panel only shows
  clusters for which uncertainties in the photometric measurements are smaller
  than 0.4 magnitudes, whereas the bottom panel only displays the clusters with
  larger photometric uncertainties.  As a result, we show that most of the
  clusters bluer than the region covered by the models have significantly larger
  color uncertainties. Clusters are corrected for reddening using their best fit
  values.
  }
  \label{fig:colorage}
\end{figure}

Figure~\ref{fig:colorage} shows the inferred dereddened optical colors of
the clusters in the sample. These colors are generated by dereddening the
clusters according to their best extinction estimate at their best age (and
mass) estimates. This two panel figure distinguishes clusters with typical
photometric errors (top panel) from the ones with abnormally large
uncertainties, due primarily to uncertainties in background determinations. In
both plots we indicate the color-age space covered by the models in blue. The
comparison between observations and best fit values from our models confirm the
self-consistency of the analysis: clusters find logical estimates when
uncertainties are typical (top panel), while large uncertainties may result in
unsatisfactory estimates (bottom panel).

\subsection{Age distribution of the clusters} 
\label{sec:yr1properties_ages}

The age probability distribution represents the apparent age distribution of the
clusters, resulting from the combined underlying cluster formation history, the
cluster destruction rate, and  observational selection effects.  The distinction
between the observed age distribution and the history of the cluster formation
rate is analogous to the difference between the present day and the initial mass
function.

We can derive the age distribution of the ensemble cluster population. {In
principle, we derive this distribution by optimizing our prior age distribution
leading to a more complex framework that we will further develop in the context
of cluster disruption. In the context of this present work, we approximate this
distribution with the co-add of the age probability distribution functions
(PDFs) of each individual cluster}.  This procedure tends to increase the
presence of tails in the distributions, however it produces a more accurate
global age distribution than assigning each cluster to a single ``best-fit''
age, especially in the light of complex and often multi-modal PDFs. The
combination of PDFs also preserves the relative quality of the fits between
clusters. Moreover, probability distributions are more robust to binning
effects.

\begin{figure}
  \centering
    \includegraphics[width=\columnwidth]{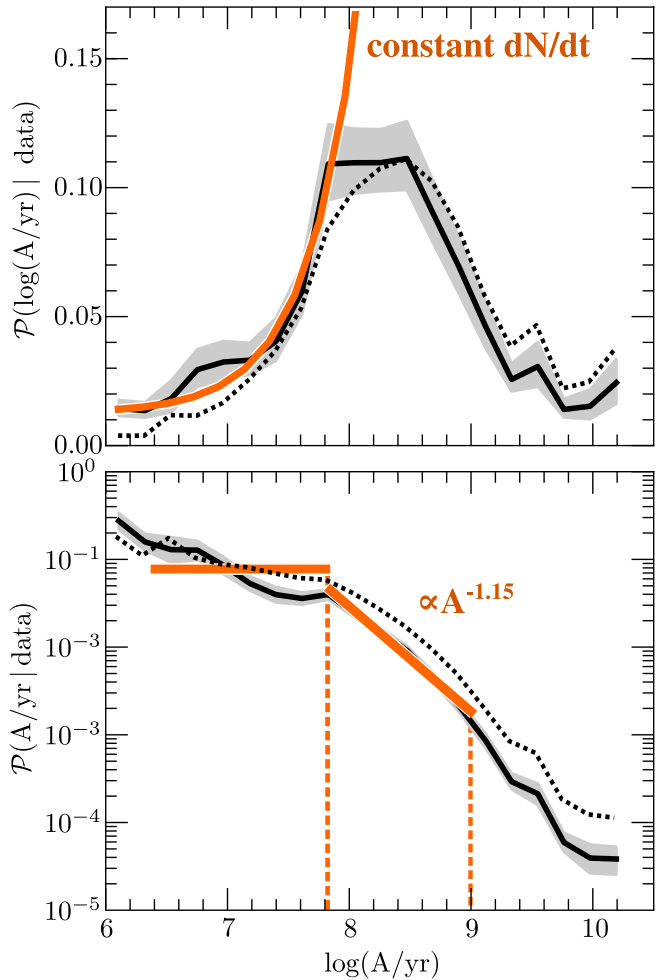}
  \caption{Age distribution (marginalized over mass and extinction) of the
  entire sample regardless of the positions of the clusters.  Thick black lines
  represent the total distribution while shaded regions are their respective
  uncertainties based on bootstrapping the cluster sample as described in the
  text (\S \ref{sec:yr1properties_ages}).  The orange thick line on the top
  panel represents a constant number of clusters, over the last $100\Myr$. It
  corresponds to the horizontal line on the bottom panel.  To the latter, we
  have also added the best fit described in \S \ref{sec:yr1properties_ages}.
  Also shown of both panels, a dotted line corresponds to the age distribution
  restricted to masses above $10^3\msun$, where effects of observational
  completeness are limited.  }
  \label{fig:marginal_age_dist}
\end{figure}

The composite distribution of ages for the ensemble of clusters is shown in
Fig.\,\ref{fig:marginal_age_dist} (black solid line). This figure shows two
representations of the age distribution of the ensemble of clusters using both
PDF representations as a function of logarithmic age.  These probability
distributions show the relative number of clusters (not mass in clusters) as a
function of age.  The top panel shows the probability distribution of
logarithmic ages, $dN/dlog(A)$, in contrast with the bottom panel showing the
distribution of age, $dN/dA$; $\mathcal{P}(\log(A))$ is defined such that the
average number of stars in a given logarithmic interval $[\log(A),\,\log(A) +
d\log(A)]$ is $dN = N\, \prob(\log(A))\, d\log(A)$, while $\mathcal{\prob(A)}$
is defined such that the average number of stars found in the linear age
interval $[A,\,A + dA]$ is $dN = N\, \prob(A)\, dA$.  Hence an equal number of
clusters at every age would appear as a constant horizontal line in the bottom
panel and as an exponential function in the top one.  In both representations we
also include dotted lines showing the composite distributions when restricted to
masses above $10^3\msun$, the estimated completeness limit of our sample
over at least a few $100\Myr$.

The uncertainties in the age distribution (represented by the shaded region on
Fig.\,\ref{fig:marginal_age_dist}) are dominated by the random sampling of a
finite number of a relatively small ($\sim600$) sample of clusters. At old ages
(\eg, $5\Gyr$) where there are few clusters, including or removing a single
cluster will lead to large variations. In contrast, such an alteration of the
sample at $100\Myr$ will have less influence.  To estimate these sampling
uncertainties, we characterize the variations of the posterior distributions by
bootstrapping \citep{Efron1987, Rubin1981}.  Specifically, we make $1000$
realizations of the cluster sample, randomly drawing $601$ clusters for each but
allowing duplications, and re-derive the ensemble age distribution for each
realization.  The shaded region indicates the range containing 95\% of the
realizations of the age posterior distribution.  Each individual realization is
also used to further assess uncertainties when comparing different cluster age
distribution models. 

The distribution in Fig.\,\ref{fig:marginal_age_dist} shows that the present day
ages of clusters span from a few$\Myr$ up to $10\Gyr$.  The young cluster age
distribution ($<100\Myr$) is relatively flat. Although we note a small
decay, the age distribution is statistically consistent with a uniform
distribution over the last $100\Myr$, as indicated by the thick orange line.  If
the birth rate of clusters has been relatively steady over this same interval,
then the nearly constant age distribution suggests that the cluster disruption
processes are likely to be inefficient over $\sim100\Myr$ timescales. We revisit
this point fully in \S \ref{sec:CFH}.  If we apply a mass cut at $10^3\msun$, at
the expected mass completeness limit, the resulting cluster age distribution
(dotted line in Fig.\,\ref{fig:marginal_age_dist}) show a better agreement with
a constant rate at younger ages and little change at older ages.

The distribution drops off at ages older than $10^8 \yr$, as expected from
cluster disruption and observational selection limits (as described in \S
\ref{sec:yr1properties_globalpic}).  We fit a power-law with spectral index
$\beta$ to the observed age distribution for ages between $10^8$ and $10^9\yr$,
using a $\chi^2$ likelihood statistics: 
\begin{equation}
 \mathcal{P}(A/yr)\propto A^{\beta}. 
  \label{eq:agedist}
\end{equation}
We find that the observed present day age distribution in this interval can be
approximated with a power-law of index $\beta = -1.15\pm0.1$.  However, note
that this power-law does not map directly into the cluster formation rate since
no cuts have been applied to ensure that the same range of cluster masses is
detectable at all ages.  The distribution of ages in the older regime is
consistent with observations of other galaxies the literature
\citep[\eg,][]{Hunter2003, deGrijs2006, Chandar2010,
Bastian2012,Fouesneau2012}, which also find power-law present day age
distributions with spectral indexes close to $-1$.  

\subsection{Mass distribution of the clusters}
\label{sec:yr1properties_mass}

\begin{figure}
  \centering
    \includegraphics[width=\columnwidth]{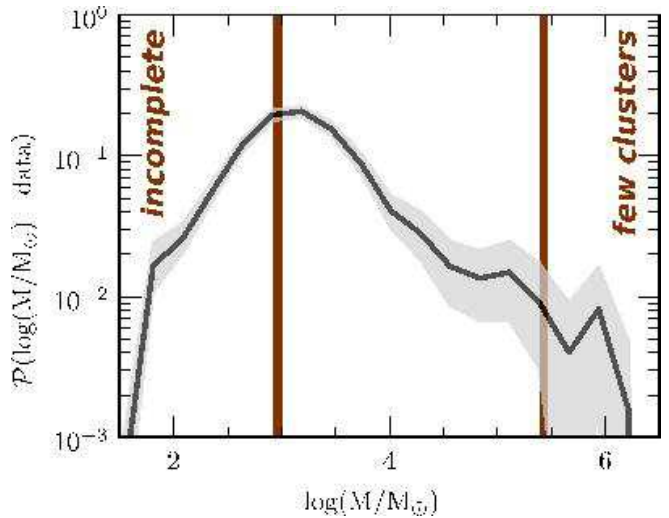}
  \caption{Mass distribution (marginalized over age and extinction) of the
  entire sample regardless of the positions of the clusters.  Thick lines the
  total distribution while shaded regions are their respective uncertainties
  based of bootstrapping the cluster sample as described in the text
  (\S\ref{sec:yr1properties_ages}).  The two vertical lines represent the regime
  limits where the distribution is less than $50$\%-complete (left) and
  stochastic presence of clusters in our sample is dominant: uncertainties are
  more than $50$\% of the value (right).  }
  \label{fig:marginal_mass_dist}
\end{figure}

The marginal distribution of cluster masses (\ie, the distribution summed over
all ages and extinctions) is shown in Fig.\,\ref{fig:marginal_mass_dist}.  We
derived the composite distribution of the whole cluster sample from their
individual probability distributions as it was done for the age distribution in
\S \ref{sec:yr1properties_ages}.  The resulting mass distribution is uncertain
in the low-mass end ($\sim10^3\msun$), where the sample is less than
50\%-complete. We adopt this limit as a lower mass limit for this distribution.
There are also significant uncertainties in the high-mass regime, where the rare
presence of one single massive cluster can induce large variations in the
distribution.  To minimize stochastic sampling of the cluster mass function, we
conservatively define a mass upper limit as the mass where the uncertainties
calculated from bootstrap resampling ({\it cf.}\,
\S\,\ref{sec:yr1properties_ages}) are more than $50$\% of the median value.  The
mass function is most reliable between these two regimes, which are delimited by
the thick vertical lines in Fig.\,\ref{fig:marginal_mass_dist}.

Over the interval of $10^3-10^{5.5}\msun$, the observed present day distribution
of masses for the entire sample can be approximated by a power-law, 
\begin{equation}
  \prob(M/\msun)\propto M^{\alpha},
  \label{eq:massdist}
\end{equation}
with index $\alpha=-1.73\pm0.11$.  Alternatively, if we restrict the sample to
only clusters with ages between $10^7 - 10^9 \yr$ in order to avoid possible
incompleteness, the distribution becomes steeper with an index of
$-1.89\pm0.12$, over the same mass range, but remains statistically compatible
with the fit of the full distribution.  Overall, we find that the present day
cluster mass function is well-described by a power-law, which agrees well with
previous analyses of other galaxies (\eg, in Fig.\,10 of the review from
\citealt{Zwart2010}). In particular we find that the younger ($<1\Gyr$) cluster
mass distribution is a close match to a power-law distribution with spectral
index $-2$, in agreements with several other cluster mass function
determinations that find indices close to $-2$ \citep[\eg][]{Zhang1999,
McCrady2007, Bik2003, Chandar2010, Popescu2012}.  Variation from the
canonical $-2$ index of the present day mass function may be the result of
stochastic formation of the cluster population, in which a variation of one
massive cluster could induce such variation. Variations may also be the result
of systematics that we explore in Section\,\ref{sec:discussion}.

The mass distribution has an apparent peak near $10^3\msun$, which falls off
towards lower masses.  Based on the Monte-Carlo collection, we know that
$\log(M/\msun) = 3.1$ is the threshold mass under which more than half of the
clusters have fluxes below the 50\%-completeness limit (defined in \S
\ref{sec:data}) in the F475W photometric passband.  This peak is therefore more
likely to be due to the incompleteness of the cluster sample at low fluxes, than
a real feature in the cluster mass distribution.  The exact characterization of
this peak is uncertain, because: (i) being faint, the low mass objects have
larger observational errors than massive ones; (ii) clusters that have low
masses while remaining above the detection limits must be young, and therefore
are in the regime most sensitive to the stochastically sampled cluster models;
and (iii) the detection of such low mass clusters becomes inefficient due to
their low contrast against the field star background. 

\subsection{Environmental Variations}
\label{sec:yr1properties_spatial}

The present day age and mass distributions discussed in \S
\ref{sec:yr1properties_ages} and \ref{sec:yr1properties_mass} were derived for
the sample as a whole. They therefore include clusters in the bulge (B01), which
has a distinct cluster population observed with a different detection
efficiency.  In this section we explore potential variations in distributions of
cluster properties at different positions within the galaxy to make a first
assessment of environmental dependencies in the cluster population.

\begin{figure*}
  \centering
    \includegraphics[width=0.9\textwidth]{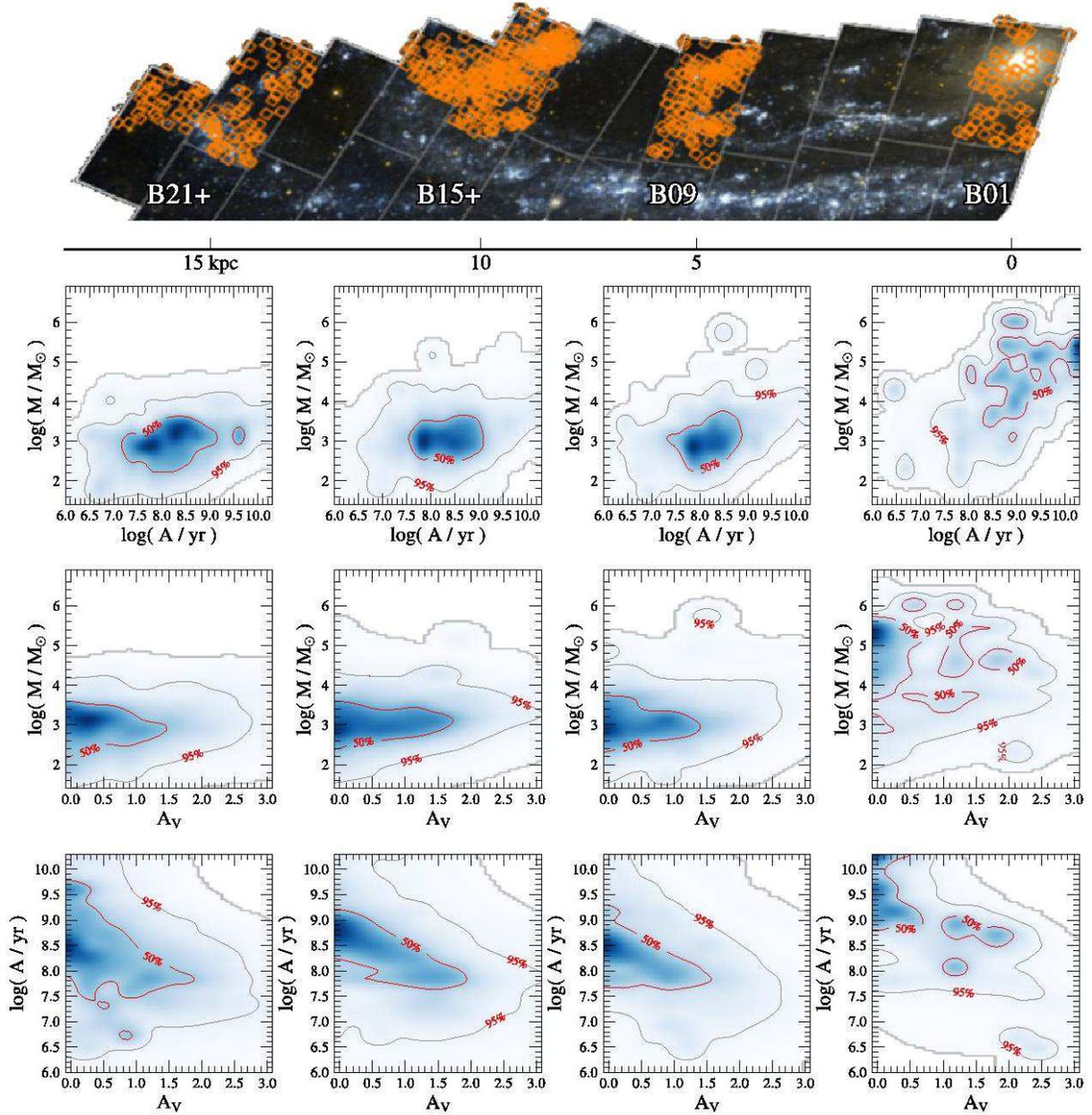}
  \caption{ Cluster mass, age and extinction distributions in different radial
  distance regimes in M31. The panels in the bottom three rows show different
  2-D projections of the joint Age-Mass-Extinction ($A-M-A_V$) posterior
  distribution.  Each column represents one of four different sub-regions of M31
  (shown on top row, with four clumps of cluster positions at B21+, B15+, B09
  and B01, left to right columns, respectively). These are analogs of
  Fig.~\ref{fig:dist}, which shows the equivalent distributions summed over the
  entire sample.  Apart the bulge, other regions exhibit similar distributions
  with possible completeness variations.
  }
  \label{fig:dist_per_brick}
\end{figure*}

\begin{figure*}
  \centering
    \includegraphics[width=0.9\textwidth]{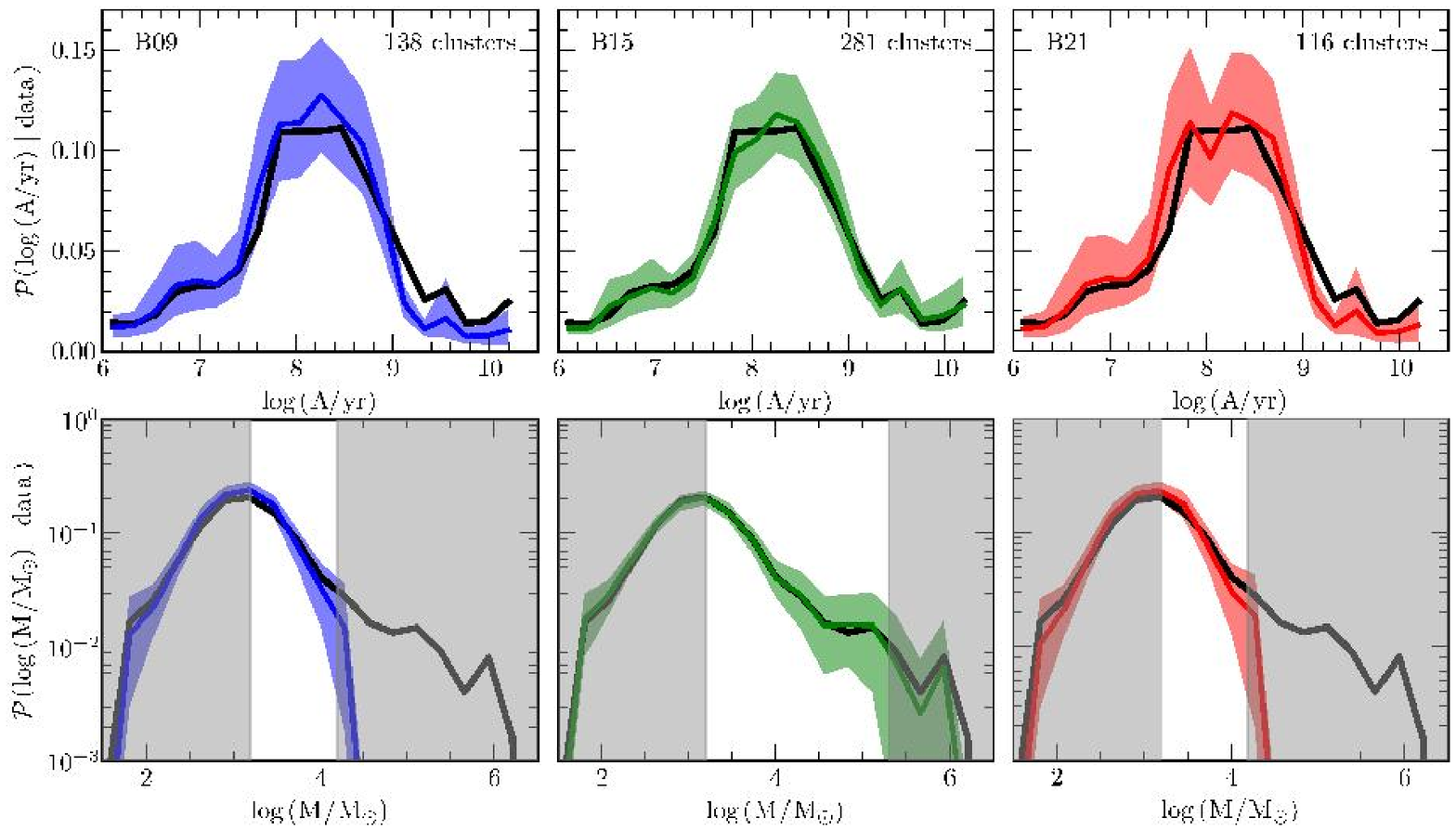}
  \caption{ 
  The individual marginalized probability distributions for age (top row) and
  mass (bottom row), for clusters in Bricks 9, 15+, and 21+ (left to right,
  respectively)
  Each column represents the distributions derived within a single brick. Shaded
  colored regions indicate the uncertainties in the distributions, derived from
  bootstrap re-sampling.
  For reference, the black lines show the distribution derived for the entire
  sample (\eg, Figs.\,\ref{fig:marginal_age_dist} \&
  \ref{fig:marginal_mass_dist}).  Gray shaded regions on the mass plots indicate
  regions where either completeness (left shading) or stochastic sampling of the
  massive clusters (right shading) are dominating the distributions. These
  regions are not included when fitting the distributions with power-laws. 
    }
  \label{fig:dist_per_brick_marg}
\end{figure*}

Figure~\ref{fig:dist_per_brick} shows the 2-D joint parameter distributions for
each of the bricks (\ie, equivalent to Figure~\ref{fig:dist}, but subdivided by
regions).  Figure~\ref{fig:dist_per_brick} highlights the variable completeness
of our sample across the galaxy; indeed, it is very difficult to detect low-mass
and/or highly extinguished clusters in the bulge, because of the high luminosity
background.  In contrast, it becomes straightforward to find them in the outer 
the disk.  From the four top panels of Fig.~\ref{fig:dist_per_brick}, one can
see that the effective 95\% completeness moves $\sim1$ dex in both age and mass,
such that older and lower-mass objects are more easily detected at larger radii.
This effect is reflected in the translation to the left of the diagonal limit in
the bottom right corner of the upper row of the plots.

Figure~\ref{fig:dist_per_brick} shows that the difference between the bulge and
the other regions is very strong. Most of the massive old
clusters are in the bulge, making up the bulk of the globular clusters
from our sample. In contrast, the lowest-mass clusters are mainly in
the outer regions of M31 (B15 and B21), which have the lowest stellar background
density, and thus the best contrast for detecting low-luminosity clusters.
Moreover, all of the disk fields are quite similar, beyond the small variations
in sensitivity.  As a result, for the rest of this paper will consider only the
disk fields (B09, B15, and B21).

Figure~\ref{fig:dist_per_brick_marg} compares the marginalized age and mass
distributions for each of the three disk regions, following the same conventions
as Figs.\,\ref{fig:marginal_age_dist} and \ref{fig:marginal_mass_dist}.  The
black line reproduces the distribution of the whole sample, for reference. There
are roughly twice as many clusters in B15 relative to the two other regions,
leading to larger uncertainties in the inner (B09) and outer (B21) regions.

\paragraph{Age distributions}

\begin{figure}[ht]
  \centering
    \includegraphics[width=\columnwidth]{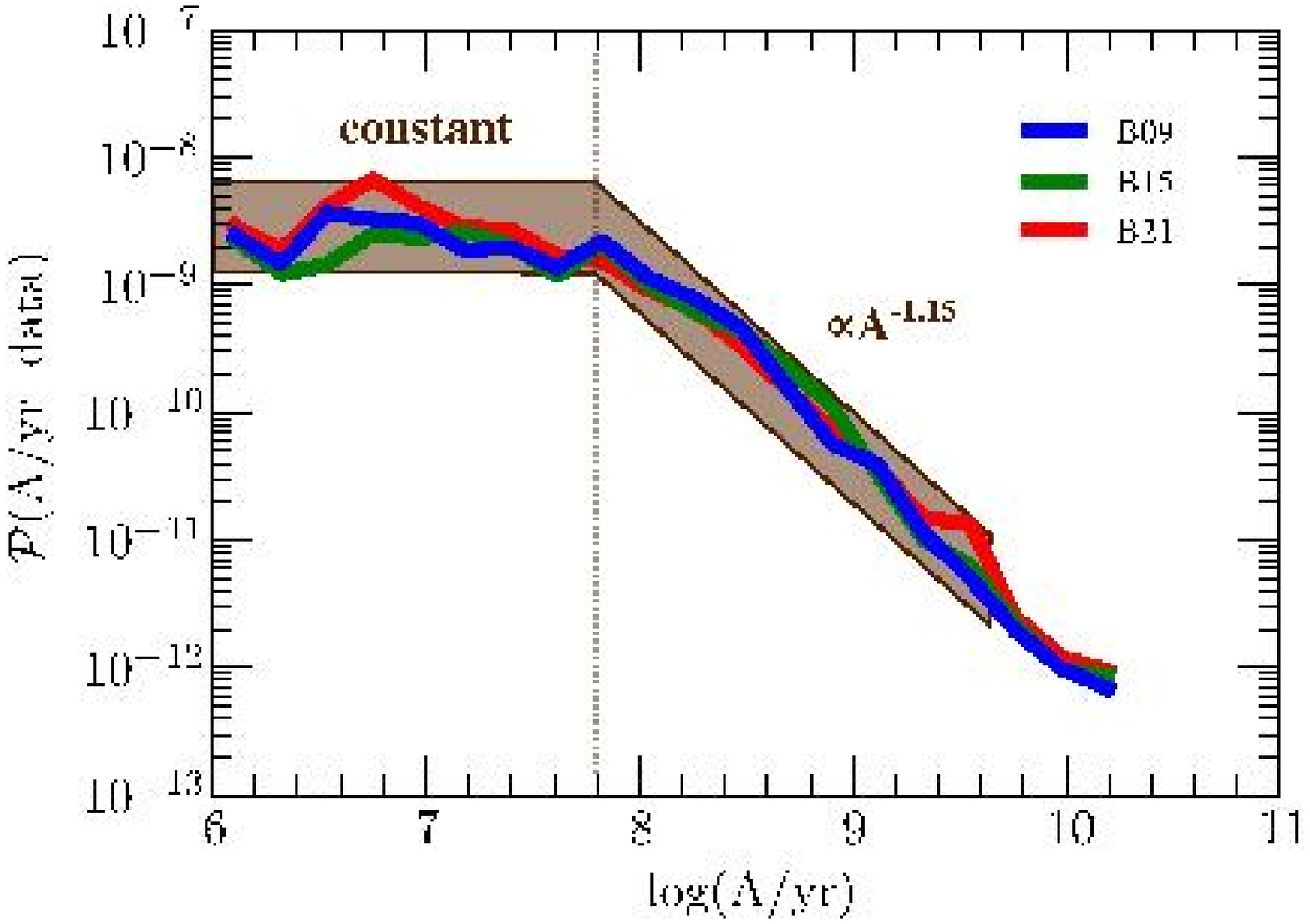}
  \caption{ 
  Distributions of cluster ages per linear age bins as a function of log(age)
  for each individual brick. This corresponds to the bottom panel of
  Fig.\,\ref{fig:marginal_age_dist} for the clusters in each of the three
  regions in the disk.  The dark shaded region illustrates the cluster age
  distribution described in \S \ref{sec:yr1properties_spatial}: constant over
  $100\Myr$ followed by a power-law decline with index $-1.15$.
    }
  \label{fig:dndt_dist_per_brick}
\end{figure}

The top panel of Figure~\ref{fig:dist_per_brick_marg} shows that there are no
significant radial changes in the local age distributions, when average over the
scale of the PHAT bricks ($1.5\times3\kpc$ at the distance of M31). We can see
this similarity more clearly in Fig.~\ref{fig:dndt_dist_per_brick}, where we
overlay the present day age distribution for all 3 disk regions. 

At young ages ($< 100\Myr$), all three age distributions are consistent with a
uniform distribution over the last $100\Myr$ suggesting that any variations in
the cluster formation rate were coherent over the galaxy.

At older ages ($>100\Myr$), the age distributions in B09, B15, and B21 are all
compatible with a power-law.  The power-law spectral indices of the age
distributions from $10^8$ to $10^9\yr$ in Fig.\,\ref{fig:dist_per_brick_marg}
are $-1.39\pm0.1$,$-1.11\pm 0.1$, and $-1.21\pm0.1$, for Bricks 9, 15, and 21,
respectively, which can be compared to the power-law of index $-1.15\pm0.1$
derived for the whole sample (\S \ref{sec:yr1properties_ages}).  The broad
similarity among the age distributions (illustrated in
Figure~\ref{fig:dndt_dist_per_brick} by the orange shaded region) suggests a
common cluster formation history within the three different forming regions. In
other words, the net result of changes in the cluster formation rate and
destruction rate were coherent across the galaxy, in spite of the fact that the
disk cluster sample spans a broad range of environments. In particular, we would
have expected that variations in the local gas density should lead to different
cluster disruption efficiencies \citep{Boutloukos2003, Lamers2005}.  However we
do not observe statistically significant differences in the age distributions of
clusters from the ring (B15) and the most outer-region (B21), where the gas
density is the highest and the lowest, respectively.  Note that this spatially
resolved analysis suggests that the slight deviation from consistency with
Fig.\,\ref{fig:marginal_age_dist} was primarily due to the inclusion of clusters
in the bulge field.

Unfortunately, the current data are not sufficient to distinguish statistically
significant variations from brick to brick. While there is a hint of a radial
variation in the position of the roll-over of the distribution, a reliable
interpretation of the roll-over is difficult without larger samples and a better
characterization of the effects of the completeness of the sample. We therefore
postpone this analysis to the full PHAT dataset.

\paragraph{Mass distributions}

\begin{figure}
  \centering
    \includegraphics[width=\columnwidth]{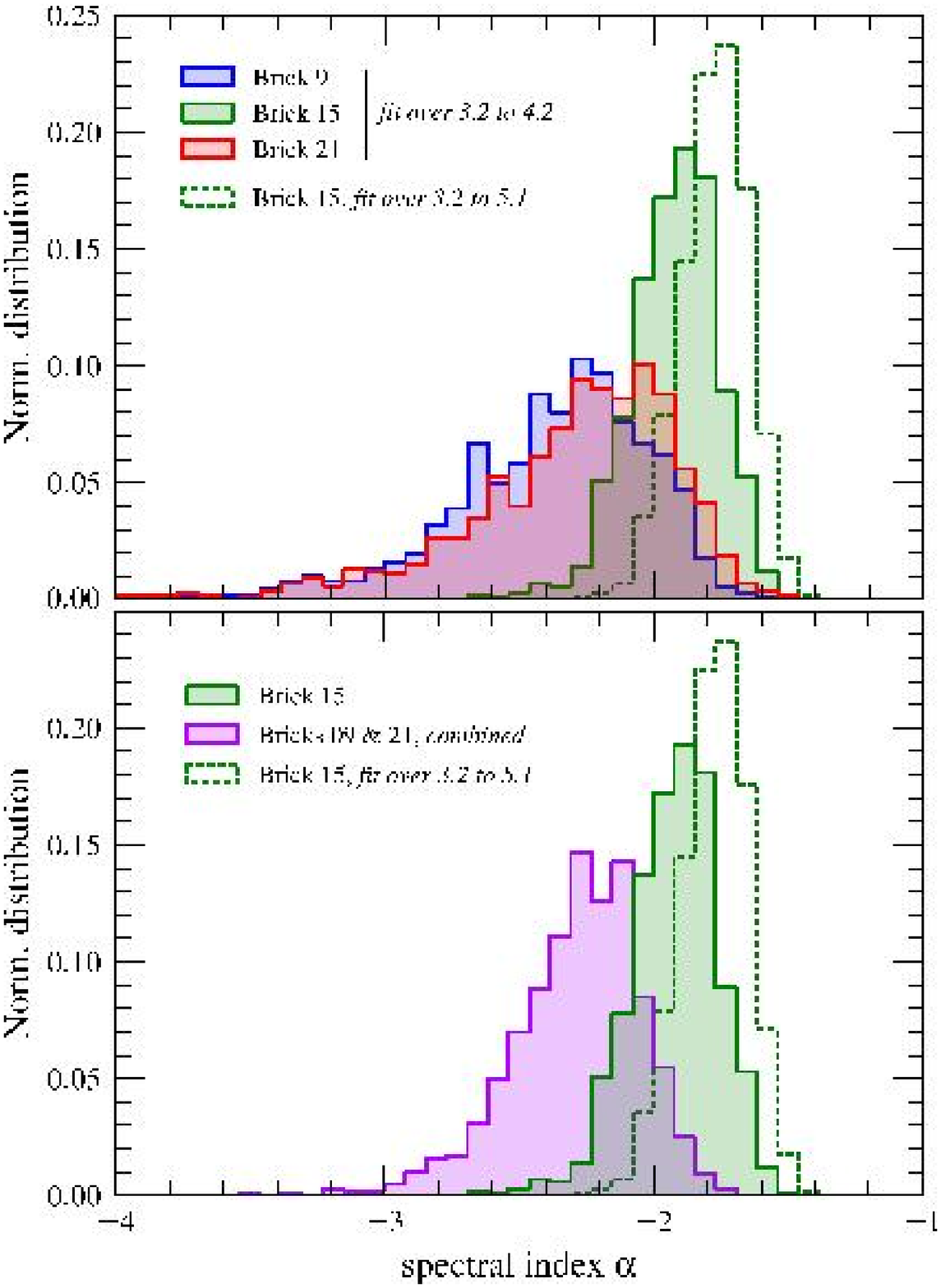}
  \caption{ 
  The distributions of the spectral indexes recovered from
  fitting a power-law to the mass distributions of the 3 outer regions
  in Fig.~\ref{fig:dist_per_brick_marg}. 
  On the top panel, we show the distributions recovered from fitting over the
  same mass range: $3.2 \leq \log(M/\msun) \leq 4.2$.  Because they appear
  similar, on the bottom panel, we grouped Brick 9 and 21 during the fit in
  order to increase the number of clusters.  
  In both panels, the dotted histogram represents the distribution
  recovered if one fit the mass distribution of Brick 15 (``10 kpc-ring'') up to
  $\log(M/\msun) = 5.1$, limit at which uncertainties dominate the mass distribution.
    }
  \label{fig:spectral_dist_per_brick_marg}
\end{figure}

The bottom panels of Fig.\,\ref{fig:dist_per_brick_marg} show the mass
distributions of all clusters in the 3 disk fields.  The shaded regions follow
the same conventions as in Fig.\,\ref{fig:marginal_mass_dist}.  The lower-mass
regime is defined to be the region where the sample is less than $50$\% complete
(\ie, $M < 10^{3.3}\msun$). By excluding this mass region, we remove sensitivity
and ensure that we have the same high completeness in all three disk regions.
We also set an upper mass limit to be where the uncertainties are more than
$50$\% of the median value (as explained in \S \ref{sec:yr1properties_mass}).
The upper mass limit is very different in the $10\,\kpc$ star forming ring (B15)
due to the presence of more massive clusters than in the two other regions (B09
and B21).  This difference is primarily due to the factor of 2 to 3 larger in
the number of clusters in Brick 15.  For stochastic sampling of the
low-frequency tail of a power-law distribution, the number of massive clusters
in each region is expected to vary much more than this factor of $2-3$, and
instead should vary by up to a factor of 10 (see Table\,\ref{tab:sumval}). In
other words, while B15 has a greater number of massive clusters, it still has
fewer than we would expect to follow the same mass function. We further discuss
the lack of massive clusters in \S \ref{sec:stoSampling}.

Qualitatively, B09 and B21 share a similar present day mass distribution within
the mass range where the estimated distributions are reliable. At first glance,
however, the mass distribution in B15 appears to be different, although this
difference may partially reflect the much larger mass range that can be probed
reliably in Brick 15.  

To quantify the differences among the bricks, we have fit each distribution with
a power-law distribution. Based upon the bottom row of
Fig.\,\ref{fig:dist_per_brick_marg}, a fit to the mass distribution of Brick 15
may be more consistent with the two other regions if we limit the mass interval
to a common range. Therefore, we restrict the fits to both the reliable mass
ranges ({\it c.f.}\,\S\,\ref{sec:yr1properties_mass}) and a common mass range
($3.2 < \log(M/\msun) < 4.2$), while accounting for their associated
uncertainties.  We repeat this exercise using bootstrap resampling of the
cluster sample in each brick.  The top panel of
Fig.\,\ref{fig:spectral_dist_per_brick_marg} shows the resulting mass spectral
index probability distributions for each individual brick.  We obtain broad
distributions for $\alpha$ in B09 and B21, as expected from the limited mass
range used during the fit and the smaller number of clusters.  The dashed green
line shows the fit of Brick 15 mass distribution when including the full
reliable mass interval, while the solid version shows the resulting distribution
when restricting B15 to the same mass range as B09 and B21, of $3.2 <
\log(M/\msun) < 4.2$.  We find a relatively narrow distribution for B15 given
the large number of clusters, and as expected, limiting the mass interval favors
steeper mass functions by $\sim0.2$\,dex.  When fitting over the common mass
range, we obtain the following values and standard deviations: B09:
$-2.2\pm0.17$, B15: $-1.8\pm0.1$, and B21:$-2.1\pm0.14$ ($-2.24\pm0.1$ for the
full reliable interval).  Although all of the three distributions of $\alpha$
are compatible (within $2$-$\sigma$) with a single power-law of index of $-1.8$
(found in \S \ref{sec:yr1properties_mass}), the differences from this overall
description are close to $3$-$\sigma$.  

Given both B09 and B21 share a similar environment and that they also seem to
follow a similar age-mass distribution, we combined those
samples into a super region probing spiral arms. This combination increases the
number of clusters in the statistics allowing us to compare 2 distinct types of
star forming regions in M31: spiral arms and the ring.
The bottom panel of Fig.\,\ref{fig:spectral_dist_per_brick_marg} compares the
distributions for B15 from the top panel to the distribution obtained for
the combination of B09 and B21 cluster samples. As expected, the combination of the two
regions narrows down the spectral index dispersion. However the resulting
distribution does not reconcile with what we find in B15 (best fit of
$-1.8\pm0.12$), and the resulting discrepancy is statistically increased. As the
PHAT survey will eventually cover inter-regions and include many more clusters,
we will be able to fully characterize the mass variations with environmental
conditions.

\begin{table*}
  \centering
  \caption{Field properties}
  \begin{tabular}{rccccccc}
    \hline\hline
    Brick        & $R_{gal}$ & F475W$_{lim}$& $N_{cl}$ & $N_{cl}$      & best          & best           & $M_{max}$          \\
    Name         &  (\kpc)   & at 50\%      &          &($M>10^3\msun$)& $\beta$       & $\alpha$       & $\msun$         \\
                 &           & (1)          &          &               & (2)           & (3)            & (4)             \\
    \hline                                                                   
    B01          & $0   $    & -3.8         & 61       & 61            & --             & --            & --              \\
    B09          & $6   $    & -3.5         & 138      & 70            & $-1.11\pm0.1$  & $-2.2\pm0.17$ & $2.5\times10^4$ \\
    B15          & $10  $    & -2.8         & 281      & 165           & $-1.21\pm 0.1$ & $-1.8\pm 0.1$ & $1.7\times10^5$ \\
    B21          & $15  $    & -2.2         & 116      & 54            & $-1.39\pm0.1$  & $-2.1\pm0.14$ & $2.0\times10^4$ \\
    \hline
  \end{tabular}
  {
  \begin{flushleft}
    \small
    (1) 50\% Completeness limit in F475W \citep{Johnson2012a},\\
    (2) best age  spectral index: $\prob(A/\yr)\propto A^{-\beta}$, with $8 < \log(A/\yr) \leq 9$ (\S\ref{sec:yr1properties_spatial}),\\
    (3) best mass spectral index $\prob(M/\msun)\propto M^{-\alpha}$, with $3.2 < \log(M/\msun) \leq 4.2$ (\S\ref{sec:yr1properties_spatial}) ,\\
    (4) maximum expected mass from a power-law with index given by their best
    $\beta$ (\S\ref{sec:systematics}). 
  \end{flushleft}
  }
  \label{tab:sumval}
\end{table*}

\section{Discussion}
\label{sec:discussion}

\subsection{Exploration of possible systematics}
\label{sec:systematics}
We have compared the cluster populations in 3 star forming regions in M31, and
may have identified the first clear evidence of variation of the present day
cluster mass function within one galaxy. On the other hand, we find very
similar age distributions in the same three regions.  We now explore the
possible observational artifacts that could be affecting our analysis.

\paragraph{Systematics from the analysis method}
We have made multiple assumptions when deriving ages and masses for the cluster
samples. Some assumptions could potentially lead to systematic errors in the age
and mass determinations. 

First, we assumed in the models that no ionizing photons escape from the cluster.
If some photons do escape, then we will over-estimate the flux in the
nebular emission (continuum + lines), and thus, the model colors will vary. These
color variations will shift clusters to different apparent ages, with little
change in the inferred masses. Moreover, these effects are in the opposite
sense of what is needed to explain the data where we find very similar age
distributions, but differences in the observed mass functions.

Second, there may be some level of inconsistency between the stellar evolution
models and the actual clusters, which could lead to biases in the derived ages
and masses. However, such an effect would be apparent in all regions, and would
not produce radial variations.

Third, the lower-mass limit of the current collection is sufficiently low that
the choice of the stellar IMF or its sampling method can affect the derived
SEDs. If real clusters have a stellar mass function different from that assumed,
then the cluster ages and masses would be biased. Again this would not produce a
radial variation in the mass distribution of the clusters, unless
the stellar IMF were also environmentally dependent.

Finally, our choice of priors during the analysis of cluster colors may not be
optimal.  We assumed uniform expectation in age and mass on logarithmic scales,
but our tests have shown that varying the prior assumptions does not
significantly affect the resulting distributions.  Therefore, our choice of prior 
is unlikely to to produce variations in the cluster mass distribution across the
galaxy. 

Apart from an environmental variation of the initial stellar mass
function, none of the above possibilities appears likely to produce the
observed radial variations in the cluster mass function while keeping similar
age distributions. All seem likely to affect all of the regions in a similar
manner (outside of the bulge).

\paragraph{Variations in the fraction of bound clusters} Our analysis is based
on the cluster sample from the \citet{Johnson2012a} catalog. Like any catalog,
the resulting sample has biases that reflect how the clusters were selected.
\citet{Johnson2012a} adopted a definition of a ``star cluster'' to be a group of
stars assumed to form a coeval population (\ie, single age, metallicity, etc.).
This definition includes any clustered stars regardless of whether they are
gravitationally bound or not. It is therefore possible that the observed
variations are due to radial changes in the relative numbers of gravitationally
bound clusters and unbound associations. Associations are most likely young
because of their intrinsic instability \citep{Gieles2011a}. When their size is
comparable to stellar clusters, they are also likely to be relatively low mass.
Therefore, increasing the proportion of associations will lead to a higher
fraction of clusters with young ages and low masses.

However, we do not observe a significant change at young ages between the three
different regions.  Including many low-mass objects could result in a steeper
mass function.  If there is a systematic radial change in the fraction of
associations, we expect this fraction to be the highest in B21, which is the
ideal environment for finding faint objects because of its low background and
relatively low density of sources and extinction.
However, as shown in \S \ref{sec:yr1properties_spatial}, the present day mass
function of B21 is similar to that of B09, which is the inner most region used
in this comparison. When combined with the fact that we see no obvious
variations in the observed age distributions at recent times, it seems highly
unlikely that variations in the fraction of bound clusters are systematically
affecting the observed age and mass distributions.

\paragraph{Completeness variations and missing low-mass clusters}
The completeness of the PHAT cluster sample is a complex function of 6 filters
photometry affected by measurement errors, which thus does not translate into a
single mass cut in the age-mass plane.  Moreover, clusters in this sample are
partially resolved into stars, which increases the complexity of estimating the
completeness accurately.

From the observational limits given in \citet{Johnson2012a}, we derived
an approximative mass completeness of $\sim10^3\msun$ using our collection of
synthetic models.  

Instead of deriving a complex completeness function, we instead applied a
conservative mass cut of $10^{3.2}\msun$, based on combining the observational
limits given in \citet{Johnson2012a} with our collection of synthetic models.
Above this mass, we expect the sample to be essentially $100$\% complete, except
perhaps at the very highest extinctions. This expectation is born out by the
distributions in Fig.\,\ref{fig:dist_per_brick}, which show no obvious signs of
incompleteness at the $10^{3.2}\msun$ level.

Brick 15 is the most gas rich of the 3 star forming regions. It is possible that B15
contains more dust, and this results in higher incompleteness due to dust
extinction.  We can estimate the number of potentially ``missing'' clusters in
Brick 15 by deriving samples to make an intrinsic mass power-law with index
$\alpha = -2$ appear to have a distribution of $\alpha=-1.8$ (as the median
value in Brick 15) We find that Brick 15 would need to have a factor of $\sim5$
more clusters between $10^3$ and $10^5 \msun$ to recover a spectral index of
$-2$. This corresponds to having $20$ missing clusters above $10^4\msun$, that
were undetected because of more than $2$ magnitudes of extinction.  
We find this possibility to be unlikely, given that the $A_V$ distribution of
detected clusters falls off steadily towards high extinctions. Moreover, we have
visually inspected all F160W images for embedded clusters at the locations of
dense molecular cloud from high resolution CARMA maps of B15 (Schruba et
al., {\em in prep}) and we find no evidence for highly embedded massive
clusters.

\subsection{Age distribution and the cluster formation history}
\label{sec:CFH}

In this study, we find that the present day cluster age distribution, $dN/dt$
(bottom panel of Fig.\,\ref{fig:marginal_age_dist}), is globally flat over the
last $100\Myr$, particularly where confined to the disk fields
(Fig.\,\ref{fig:dndt_dist_per_brick}).
There are other cluster samples from the literature that show the same constant
distribution at young ages (\eg, \citealt{Lamers2005}, limited to $600\pc$ from
the Solar neighbourhood;\ \citealt{Hodge1987, Chiosi2006}, flat over $1\Gyr$ in
the SMC). After this initial $100\Myr$ period, we find a power-law drop-off at
older ages consistent with index $-1.15$. This index may change when we
eventually include the complete sample from the PHAT survey, which will provide
greater weighter of regions that lack recent star formation than the Year 1
\citet{Johnson2012a} catalog.

The interpretation of the observed $dN/dt$ distributions is not straightforward,
because these distributions are heavily dependent on the observational
completeness of the sample, the birth rate of the clusters as a function of
time, and the efficiency of the cluster disruption processes as a function of
time, mass, and environment. We are therefore deferring a full analysis of the
age distribution until a subsequent paper, when we will have a larger number of
clusters in the sample and a better characterization of its completeness.

As an intermediate step, however, we can draw an initial comparison between the
observed present day age distribution and that expected under a few assumptions.
If one assumes that the cluster population formed at a constant rate and with a
{\it power-law initial cluster mass function} (ICMF), and {\it no dissolution},
\citet{Gieles2008} demonstrate that the age distribution can be analytically
estimated.  Based on continuous population synthesis flux predictions for
cluster fading, they show that the observed age distribution should follow a
power-law distribution with an index of $\sim-0.7$, if the sample is limited by
one optical band detection, and $dN/dt$ should be constant, if the sample is
mass limited.  

We find a uniform distribution for the first $100\Myr$.  If we assume that
the young clusters from \citet{Johnson2012a} are complete down to our
$10^{3.2}\msun$ mass limit, then we should obtain a constant distribution until
the age that the fading or disruption starts to remove clusters from the sample.  
 
We find that the drop-off occurs at $\sim100\Myr$, independent of the
region of the galaxy, suggesting that cluster disruption is little to no effect
prior to this timescale. In addition, the $100\Myr$ extent of the flat $dN/dt$
distribution appears to the same in all regions, suggesting that the
environmental dependence of the cluster disruption must be weak.

At older ages ($>100\Myr$) we find a power-law decrease in the observed $dN/dt$
distribution, with an index of $\beta=-1.15$. This index is steeper than the
predicted value by the cluster fading model of \citet{Gieles2008} in the
presence of a magnitude limit. This may suggest that there are some cluster
disruption effects at work above the $10^{3.2}\msun$ mass limit of our analysis. On
the other hand, the role of selection effects has not yet been fully qualified
and can potentially lead to the steepening that is observed. We will work on this
issue more fully in an upcoming paper.

\subsection{Truncated Mass Function}
\label{sec:stoSampling}
The reliable mass range over which we can fit the present day mass distribution with a
power-law functino varies from one region to another. We adopted different
limits motivated by the lack of massive clusters in B09 and B21 in contrast with
B15.  This difference could simply reflect the smaller number of clusters
overall in B09 and B21, or it could reflect that the cluster mass function in
these bricks is ``truncated'', given that other studies have found evidence for
a truncation of this power-law at the high-mass end \citep[\eg,][in M31 for the
latter]{Bastian2008, Larsen2009, Vansevicius2009}. To test for a possible
truncation, we can estimate the most-massive cluster we would expect from an
untruncated power-law mass function and compare to our observations.

To do this test, we assume that the cluster formation rate is constant and that
the cluster mass function follows a power-law mass distribution over a mass
range $[M_1, M_2]$,
\begin{eqnarray}
  \mathcal{P}(M/\msun) &\propto& M^{\alpha},
\end{eqnarray} 
where $\alpha$ the spectral index of the cluster mass function. With these
assumptions, we can define the probability of obtaining a cluster with a mass
$m$ above a given mass $M$ as
\begin{eqnarray}
  \mathcal{P}( m \geq M \mid \alpha  ) & = & \frac{\int_{M}^{M_2}{ x^\alpha\,dx}}{\int_{M_1}^{M_2}{ x^\alpha\,dx}},
\end{eqnarray} 
We need then to introduce the total number of clusters, $N_{cl}$, formed from
the mass distribution. If we consider independent draws from this mass
distribution, then the expected maximum mass $M_{max}$ satisfies the
condition that we have one and only one cluster for $m=M_{max}$,
therefore:
 \begin{eqnarray}
   1 &=& N_{cl} \times \prob( m > M_{max} \mid \alpha)\\
M_{max} &=& \left\{ M_2^{\alpha+1} - \frac{1}{N_{cl}} \times \left(  M_2^{\alpha+1} -
   M_1^{\alpha+1}\right) \right\} ^ {-(\alpha+1)} \label{eq:Mmax}
   \label{eq:mass_sto1}
 \end{eqnarray}
If we consider the upper mass limit $M_2$ to be infinite (\ie, no truncation),
and $\alpha < -1$ for convergence, then:
 \begin{eqnarray}
   M_{max} = {M_1} \times N_{cl} ^ {\alpha+1} 
   \label{eq:Mmax_infLim}
 \end{eqnarray}
 
\begin{figure}[t]
  \centering
    \includegraphics[width=\columnwidth]{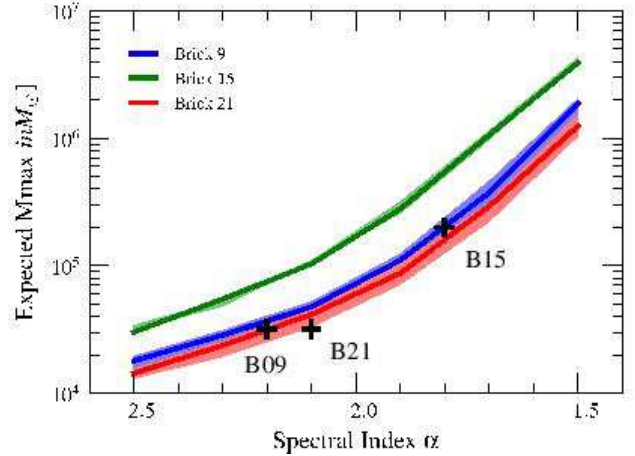}
   \caption{ Distributions of expected most massive cluster mass as a function
   of the power-law spectral index, in each of the 3 outer regions of the study.
   Shaded regions are based upon the Poisson variations of the number of
   clusters per brick.  Black crosses are the mass cut applied during the fit in
   \S\ref{sec:yr1properties_spatial} for the labeled regions. }
  \label{fig:stoSamp}
\end{figure}

Figure\,\ref{fig:stoSamp} shows the different expected distributions of
$M_{max}$ as a function of the power-law index $\alpha$, for each field brick in our study.
The different curves are
constructed from Eq.\,\ref{eq:Mmax}, assuming the mass function is defined within
$10^3-10^7\msun$, to match our typical completeness
limits. The number of clusters $N_{cl}$ above $10^3\msun$ is taken
from Table\,\ref{tab:sumval} for each region.  
We compare the predicted maximum cluster mass with the observed values derived
from the data in \S\,\ref{sec:yr1properties_spatial}. Because the masses were
derived from PDFs rather than single best fit values, our estimated $M_{max}$ is
taken to be the point where the uncertainties on the mass distribution become
more than 50\% of the estimated value. 

In Figure\,\ref{fig:stoSamp}, 
The values for each brick are indicated by the black crosses, for which the
spectral indices correspond to the best values obtained in
\S \ref{sec:yr1properties_spatial}. 

The comparison between the predicted and observed values of $M_{max}$ shows that
the lack of massive clusters in B09 and B21 is consistent with the expectation
for a stochastic sampling of a power-law mass distribution. In other words,
there is no need to invoke mass truncation to explain the observed upper mass
limits of the present day cluster mass function. The mass cut we observe in B15
($M_{max} = 1.7\times 10^5\msun$) appears to be a factor of 2 smaller than the
theoretical sampling prediction of the maximum mass ($M_{max} = 5.2\times
10^5\msun$).  There are only $3-4$ clusters between the adopted $M_{max}$ and
the theoretical prediction value, which leads to large uncertainties at this
regime. We therefore have adopted our more conservative value of $M_{max}$.

\section{Conclusions} 
\label{sec:conclusions}

We have derived ages, masses and exintction for the Year 1 PHAT cluster sample
\citep{Johnson2012a}, by comparing the cluster integrated 6-filter fluxes with
an extended version of the stochastically sampled model clusters presented in
\citet{Fouesneau2010}.
The locus of the collection of stochastic models in color space (\eg, Figure 2)
shows excellent agreement with that of the collection of cluster observations.
Clusters with broadband colors either bluer or redder than those of the
traditional continuous models find a natural match with the models we used in this paper.

We generated the full joint probability distribution function of the 
age, mass, and extinction for each of the 601 individual clusters in the sample.
We then combined their individual distributions into global cluster age and mass
distributions, noting limits at which completeness issues in the sample become
severe.

The sample of clusters spans the entire length of the age sequence and
includes a significant number of clusters with masses well below $10^3\msun$.
Only a few datasets have the ability to sample objects across a variety of
stages in cluster evolution over such a large, uninterrupted mass range.

We find that the cluster age distribution shows a constant number of clusters
over the last $\sim 100\Myr$, with a power-law decline at older ages (see
Figs\,\ref{fig:marginal_age_dist} \& \ref{fig:dndt_dist_per_brick}).  At least
above the mass of $10^{3.2}\msun$, these
results are consistent with M31 producing a constant number of clusters from
$100\Myr$ ago to present, with little significant cluster disruption over this
timescale.

The mass distribution derived from the analysis closely resembles the power-law
distributions obtained from many other galaxies.  Specifically, the overall
power-law index of the mass distribution is consistent with the canonical value
of $-$2. However, the current cluster sample suggests a possible radial
variation of this distribution across the disk, with the shallowest power-law
found in the region with the highest star formation rate.

When we study the entire PHAT survey, including lower masses and a larger
sample of fainter clusters, the improved accuracy and time resolution
achievable with the new stochastic methods will allow us to address new
questions. Future work will account for the challenging determination of
completeness and selection effects.  In particular, the expected number of
clusters in PHAT will eventually provide 5 times more clusters over a broad
range of local environments, which will open the possibility to study local
variations among cluster populations beyond our current the initial assessment
in this study.

\begin{acknowledgements}
   The authors acknowledge the efforts of the entire PHAT collaboration in this
   project. 
   Also, the authors thank the Nate Bastian from is prompt and useful referee
   report. 
   DG kindly acknowledges financial support by the German Research Foundation
   (DFG) through grant GO~1659/3-1.  Support for DRW is provided by NASA through
   Hubble Fellowship grants HST-HF-51331.01 awarded by the Space Telescope
   Science Institute.  This paper is based on observations taken with the
   NASA/ESA Hubble Space Telescope. Support for this work was provided by NASA
   through grant number HST GO-12055 from the Space Telescope Science Institute,
   which is operated by AURA, Inc., under NASA contract NAS5-26555. 
\end{acknowledgements}

\bibliographystyle{apj}
\bibliography{main}

\appendix

\section{Catalog}
\label{sec:appendix_catalog}
This Section present the catalog resulting from this study.
Only a few entries are shown as an example describing the full content available
online.

Here we present the catalog of parameter estimates derived in this study for the
\citet{Johnson2012a} star cluster catalog. A subset of the catalog is presented in
the print version of the paper, with the full catalog being online. 

In this table are given the cluster PHAT ID numbers and the coordinates of the
clusters from \citet{Johnson2012a}.
The ``best'' values are the coordinates of the ($A_V$ - age - mass) triplet that
maximizes the posterior distribution of the individual clusters. The other
values are the $i$th-percentiles of the marginalized distributions over the two
other parameters. The $16$th and $84$th percentiles are the equivalent limits of
a $1$-$\sigma$ range for a Gaussian distribution ($2.5$th and $97.5$th percentiles
are the limits of a $2$-$\sigma$ confidence interval.)

\clearpage
\begin{turnpage}
\begin{deluxetable*}{ccccccccccccccccccccc} 
\tablecolumns{18} 
\tablecaption{Age-Mass and extinction results from the analysis with discrete cluster models.\label{tab:catalog} }
\tablehead{ 
\colhead{}                            &  
\colhead{}                            &  
\colhead{}                            &  
\multicolumn{5}{c}{A$_V$}             & 
\colhead{}                    & 
\multicolumn{5}{c}{$\log(A/yr)$}      & 
\colhead{}                    & 
\multicolumn{5}{c}{$\log(M/M_\odot)$} \\
\cline{4-8} \cline{10-14} \cline{16-20} \\
\colhead{PCNUM}               & 
\colhead{RA\tablenotemark{1}} & 
\colhead{DEC\tablenotemark{1}}  & 
\colhead{best\tablenotemark{2}} & 
\colhead{p16\tablenotemark{3}}  & 
\colhead{p84}                 & 
\colhead{p2.5}                & 
\colhead{p97.5}               & 
\colhead{}                    & 
\colhead{best}                & 
\colhead{p16}                 & 
\colhead{p84}                 & 
\colhead{p2.5}                & 
\colhead{p97.5}               & 
\colhead{}                    & 
\colhead{best}                & 
\colhead{p16}                 & 
\colhead{p84}                 & 
\colhead{p2.5}                & 
\colhead{p97.5}               &
\colhead{cflag}\tablenotemark{4}
}
\startdata 

1       & 11.63827 & 42.19389 & 0.9 & 0.6 & 1.2 & 0.6 & 1.2    & & 6.75    & 6.53    &  6.96    & 6.32    & 6.96   & & 3.18    & 2.35    & 3.45    & 2.35    & 3.73  & 0\\
2       & 11.63714 & 42.20994 & 0.3 & 0   & 0.6 & 0   & 0.6    & & 6.53    & 6.32    &  6.75    & 6.32    & 7.18   & & 3.18    & 2.90    & 3.45    & 2.90    & 3.73  & 0\\
29      & 11.62827 & 42.22423 & 0   & 0   & 0.3 & 0   & 0.3    & & 8.47    & 8.25    &  8.68    & 8.25    & 8.68   & & 3.45    & 3.18    & 3.73    & 3.18    & 3.73  & 0\\
34      & 11.59456 & 42.19844 & 0.3 & 0   & 1.5 & 0   & 2.4    & & 8.68    & 8.04    &  9.11    & 7.39    & 9.33   & & 2.35    & 1.80    & 2.90    & 1.53    & 3.18  & 0\\
35      & 11.62181 & 42.20984 & 0.9 & 0.6 & 1.2 & 0.6 & 1.2    & & 6.96    & 6.53    &  7.18    & 6.53    & 7.18   & & 2.08    & 1.80    & 2.63    & 1.53    & 2.63  & 0\\
...     & ...      & ...      & ... & ... & ... & \multicolumn{10}{c}{\it Full table available as electronic supplement} & ... & ...     & ...     & ...     & \\    
1725    & 11.58850 & 42.25820 & 1.2 & 0.3 & 2.1 & 0   & 2.4    & & 7.61    & 6.75    &  8.25    & 6.10    & 8.47   & & 2.63    & 2.08    & 3.18    & 1.53    & 3.18  & 0\\
1726    & 11.60528 & 42.26594 & 1.8 & 0.9 & 2.1 & 0   & 3      & & 8.90    & 8.68    &  9.33    & 7.18    & 10.1   & & 4.01    & 3.45    & 4.28    & 3.18    & 4.56  & 1\\
1728    & 11.58719 & 42.25828 & 0   & 0   & 1.5 & 0   & 2.4    & & 8.04    & 6.10    &  8.25    & 6.10    & 8.68   & & 2.08    & 1.53    & 2.90    & 1.53    & 3.18  & 0\\

\tablenotetext{1}{ Significantly more figures are available in the electronic table}
\tablenotetext{2}{ ``best'' represents the triplet which maximize the posterior}
\tablenotetext{3}{  ``pXX'' represents the XX-th percentile of the marginalized posterior }
\tablenotetext{4}{  flag suspicious fit from visual CMD inspection (boolean value)}
\enddata
\end{deluxetable*}

\end{turnpage}

\end{document}